\documentclass[twocolumn]{aastex631}

\usepackage{amsmath}
\usepackage{soul}

\allowdisplaybreaks

\usepackage{color,ulem}
\newcommand{\y}{\color{red}}

\begin{document}

\title{Testing tidal theory using Gaia binaries: the red giant branch}

\author[0000-0001-9420-5194]{Janosz W. Dewberry}
\affiliation{Canadian Institute for Theoretical Astrophysics, 60 St. George Street, Toronto, ON M5S 3H8, Canada}
\author[0000-0003-0511-0893]{Yanqin Wu}
\affiliation{Department of Astronomy \& Astrophysics, University of Toronto, Canada}

\begin{abstract}
Tidal interaction is a major ingredient in the theory of binary evolution. Here, we study tidal circularization in binaries with red giant primaries. We compute the tidal evolution for binaries as their primary stars evolve along the red giant branch, under dissipation of dynamical tides in the convective envelope. We then compare this evolution with a sample of $\sim30,000$ red giant binaries reported by Gaia DR3. These binaries clearly  show the expected gradual advance of tidal circularization, as the primary expands. But some tension with theory remains. While our  calculations always predict a critical separation for tidal circularization at about $3-4$ times the stellar radii, binaries with less evolved giants are observed to be  circularized out to about twice as far.  They also exhibit an overly extended `cool island', a collection of circular orbits that reach a couple times beyond the  circularization limit. These discrepancies are reminiscent of, but less severe than,  the situation for main-sequence binaries. We also find that tides can spin giant stars up to rotation rates that should affect their mass-loss. Additionally, many binaries may begin mass transfer while still eccentric.
\end{abstract}


\section{Introduction}\label{sec:intro}
Binary evolution connects to numerous phenomena of current high interest. These include SNIa, merging neutron stars and black holes, blue stragglers, X-ray binaries, and CVs \citep[for a recent review, see, e.g.][]{Han2020}. Multiple physical processes control this evolution. Here, we focus on one major process: tidal interaction, and more specifically, the process of tidal circularization.

Tides in stars and gaseous planets are inherently complex fluid dynamical phenomena. 
Despite this complexity, nearly analytical (linear) models can in principle predict the orbital evolution caused by weak tides. A large body of research has been directed toward developing such models for stars that are fully radiative, fully convective, a mixture of the two, rotating, magnetized, and/or evolving \citep[see review by][and references therein]{Ogilvie2014}. 

\begin{figure*}
    \centering
\includegraphics[width=0.975\textwidth]{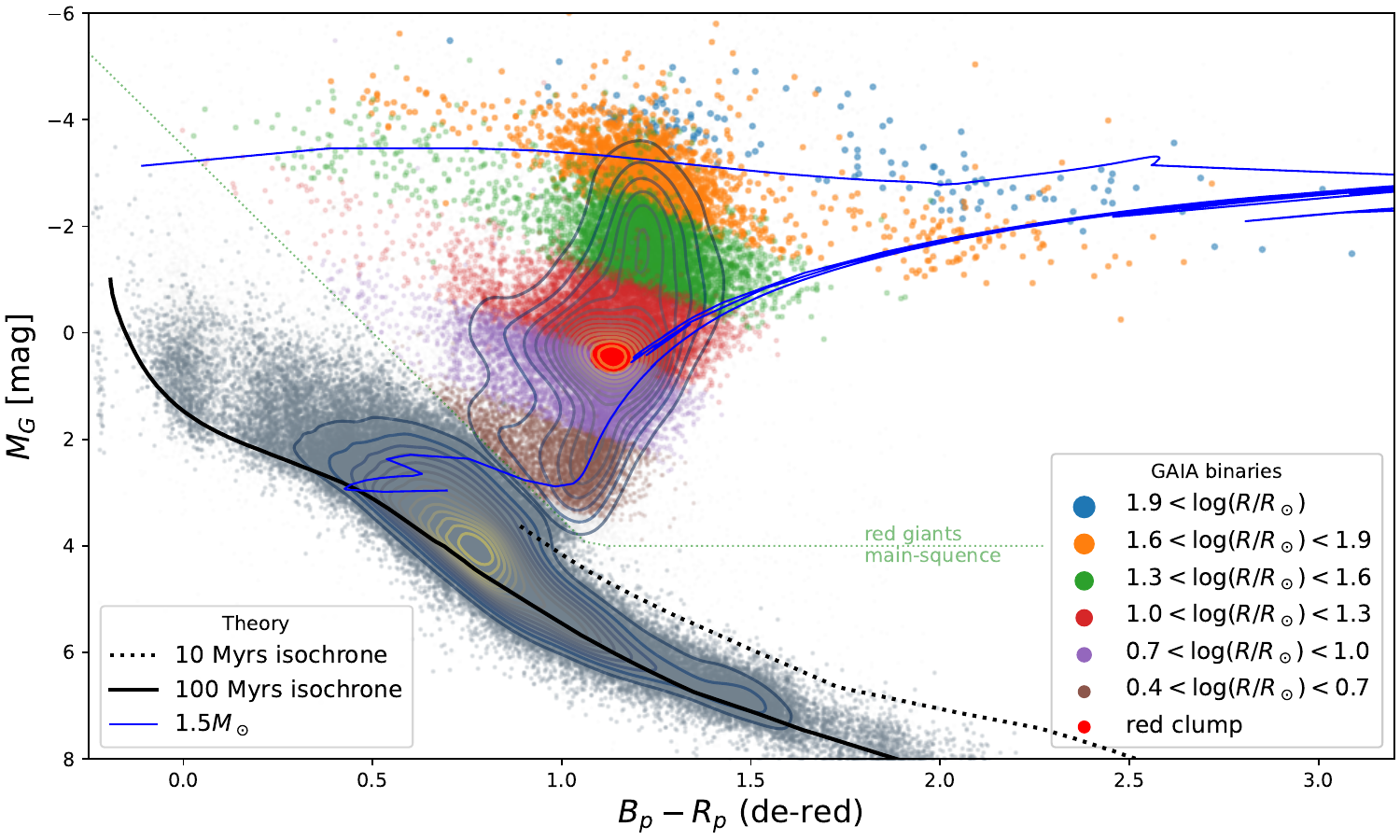}
    \caption{HR diagram for the Gaia binary sample  used in this work. The gray points indicate main-sequence binaries, while the colored denote red-giant binaries (colored by different primary radii).  The contours represent sample density. The two black curves show the isochrones for zero-age-main-sequence and 10 Myr old systems, for solar metallicity stars \citep[PARSEC isochrone,][]{parsec}. The evolutionary trajectory of a $1.5M_\odot$ star is super-imposed as a blue curve.
    }
    
    \label{fig:hrd_gaia}
\end{figure*}

The predictions of these theories have been compared against observed binaries, with varying degrees of success.
They generally fail to explain the eccentricities of late-type main-sequence binaries. 
While almost all calculations predict that such binaries should be circularized at most out to a few days \citep[e.g.,][]{Zahn1977,Terquem1998,Goodman1998,Witte1999,Ogilvie2007,Barker2022,Zanazzi2021}, spectroscopic binaries in open clusters \citep{Meibom2005,Meibom2006,Mazeh2008} and in the Gaia binary sample \citep{Bashi2023} show an abundance of circular binaries out to almost 20 days (also see Fig. \ref{fig:mainsequence}). The efficiency of tidal dissipation, as calculated from classical tidal theory, is too low by orders of magnitude
 \citep[see, e.g.,][]{Terquem1998,Goodman1998}. 

In contrast, tidal theory has enjoyed some qualified success when applied to red giants. 
\cite{Verbunt1995} pioneered the study of tidal circularization in red giant stars. They showed that a set of $28$ binaries in open clusters are consistent with the theory of \citet{Zahn1977}, where the equilibrium tidal bulge is damped by eddy viscosity in a convective envelope: $6$ of the systems predicted by equilibrium tides to be circular are indeed so. This work was extended by \citet{Price-Whelan2018} to $234$ binaries with evolved primaries from the APOGEE survey. They demonstrated that the orbital period within which all eccentricities are negligible (circularization period) increases with decreasing surface gravity (i.e., with advancing stellar evolution) in the primary star. \citet{Beck2018} and \citet{Beck2024} drew similar conclusions using giant binaries from the Kepler mission, and with seismically inferred ages (respectively).

Here, we expand on these previous studies by taking advantage of a sample of $\sim 30,000$ red-giant binaries from the Gaia DR3 catalog \citep{Arenou2023}. 
This large sample contains binaries on various stages of red giant evolution. It therefore affords us a much more nuanced---and hopefully more revealing---view of tidal circularization. By comparing these observations against detailed computations, we aim to clarify  which physics and dynamics are most essential to an accurate, predictive tidal theory. 

Here, we focus only on systems where the primaries are evolving along the red giant branch (RGB), deferring consideration of circularization on the main sequence (MS) and the asymptotic giant branch (AGB) to later work. 

Section \ref{sec:gaia} introduces the sample of binaries that we select from Gaia. Section \ref{sec:mod} then specifies the theoretical models for tides that we consider in this paper, and Section \ref{sec:res} describes the results of our numerical calculations. Finally, Section \ref{sec:obs} compares these calculations against Gaia observations, and Section \ref{sec:concl} concludes.

\section{Gaia Binary Data}\label{sec:gaia}

\subsection{Sample Extraction}

A catalogue of non-single-star sources is presented by 
\citet{Arenou2023}. We query the table {\texttt{\textbf nss$_{-}$two$_{-}$body$_{-}$orbit}} in the Gaia archive for binaries that have periods, eccentricities and lie long-ward of 2 days, obtaining $356,221$ systems. To reduce contamination, we pare down this large set using the following steps. 
\begin{itemize}
    \item We query the table {\texttt{\textbf astrophysical$_{-}$parameters}} in the Gaia archive, for systems that have extinction determinations. These have assigned astrophysical parameters such as de-reddened $B_p-R_p$, $M_g$, stellar radius, effective temperature and stellar luminosity. This reduces the catalog by about a third.
    \item We discard the $\sim 12,000$ systems that are only characterized by eclipses (`Eclipsing Binaries'). These are a subset of the $\sim 2$ million EB catalogue \citep{Mowlavi2023}.   Unfortunately, many of them have eccentricities artificially set to zero. They offer us little value. 
    \item Following the catalogue paper \citep{Arenou2023}, we retain only those with high enough significance (`significance' $> 15$) \footnote{We ascertain that our results are not affected when this threshold is set to $20$ and when the sample is $\sim 20\%$ smaller.} 
    and those short-ward of $P = 1200$days. The latter is the DR3 data span. These steps cut down the sample by about a factor of 2. 
\end{itemize}

\begin{figure}
    \centering
    \includegraphics[width=0.45\textwidth,trim=0 7 0 15,clip=]{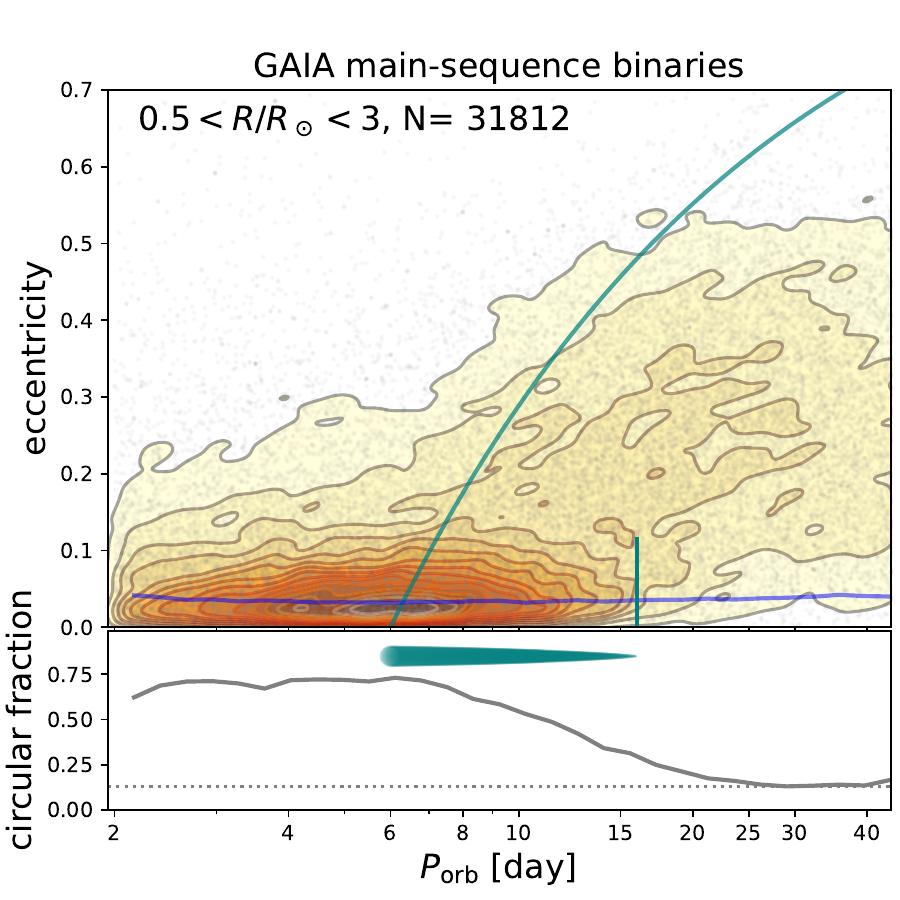}
    \caption{Eccentricity-period distributions of Gaia Main-sequence binaries, including primaries with radii from $0.3 R_\odot$ to $3R_\odot$. They are not the focus of this work and are only included here for completeness. Top panel: eccentricity-period distribution, with kernel density estimates over-plotted on the data points. Contour levels are $[0.05,0.11,0.18...1]$ and remain the same throughout this paper. The teal curve indicates orbits with  periaps equal to the separation of a 6-day circular orbit. This crudely defines the eccentricity upper envelope for all main-sequence binaries.  The blue curve indicates mean eccentricity error ($\epsilon_e$).  Bottom panel: fraction of circular binaries (defined as $e < 3 \times \epsilon_e \sim 0.1$). The dotted line  (at $\sim 13\%$) describes binaries outside 50 days and thus represents the primordial fraction. A `cool island' of circular orbits persists to much longer periods. Defining its outer edge to be  where the circular fraction falls below twice the primordial fraction, we find $16$ days (the vertical bar). The shrinking arrow encompassing (6,16) days quantifies the reach of tidal circularization. 
    }\label{fig:mainsequence}
\end{figure}

Now left with a total of $150,927$ binary systems, we further separate them by evolutionary status. As can be observed in the HR diagram of these systems (Fig. \ref{fig:hrd_gaia}), the main-sequence binaries\footnote{We will name a binary after the more luminous component, e.g., an MS binary, a giant binary.} 
can be separated from the giant binaries by the following line:
\begin{equation}
    M_G  = {\rm Min}[
        4.0, 
        7(B_p-R_p)_{\rm de-red} - 3.5
    ],
    \label{eq:L0}
\end{equation}
with those brighter being giants and those dimmer being MSs. This returns $119,425$ MS binaries and $31,502$ giant binaries.

For completeness, we present the observed eccentricity distributions for main-sequence binaries in Fig. \ref{fig:mainsequence}.  We focus on binaries with periods shorter than $30$ days. These are almost all discovered by radial velocity. The primary radii range from $0.3$ to $3 R_\odot$, with a similar spread in stellar masses. Naively, we expect the process of tidal circularization to depend most sensitively on $a/R$, the ratio between semi-major axis and primary radius. So systems with larger primaries should be circularized out to a larger $a$ and a longer period.

Surprisingly, the eccentricity distributions in period are nearly identical across all spectral types. In fact, all stars appear to obey the same upper-envelope in eccentricity-period space, one that has a peri-centre equal to the separation of a 6-day orbital period.\footnote{This corresponds to $a/R \sim 12$ for a $1.5M_\odot$ primary, and smaller for larger stars.} This upper envelope may be explained, at least for solar-type stars, by tidal theory \citep[see, e.g.,][]{Zanazzi2021}. Moreover, they all exhibit the same `cool-island' feature, circular orbits persisting well past this value to $15-20$ days, or about twice as far in separation as $6$ days. Such an extended cool island has long resisted theoretical explanations.\footnote{\citet{Bashi2023} discarded this circular population, arguing that they have small signal-to-noise ratios. However, circular orbits, by definition, have small signal-to-noise ratios. The cool island is a physically important feature in the data.}

Beyond $15-20$ days, we observe that the orbital eccentricities return to the primordial distribution, i.e., a Rayleigh distribution with a mode of $\approx 0.3$ \citep{Wu2024}. To quantify the `cool island', we set its outer limit to where the local circular fraction drops to about twice the background value (see Fig. \ref{fig:mainsequence}). This yields a period of $16$ days for all main-sequence binaries. So for main-sequence stars, [6,16] days jointly describe the limits of tidal circularization. 

\subsection{Giant Binaries}

Among the giant binaries, $78\%$ are detected by the `SB1/SB2' method, $12\%$ by  `EBSpectro' and $9\%$ by `Orbital'. This sample of binaries form the basis of our analysis in this paper. 

In Fig. \ref{fig:hrd_gaia}, the giant binaries are shown in a HR diagram. They are grouped by their respective primary radii into 6 divisions. Such a division is convenient. Not only do these radii reflect the stars' evolutionary state along the RGB, they also, as we argue below, provide suitable rulers for measuring tidal circularization.

Admixed with the red giants, there are red clump stars (horizontal branch) and AGB stars. We extract a `red clump' sample, comprising primaries that have descended the RGB and are now in stable core helium burning. These stars can be clearly seen in  Fig. \ref{fig:hrd_gaia} as the  concentration near the middle of the RGB branch. We first identify the highest density peak of the giants. This lies at $(B_p - R_P)_{\rm de-red} \approx 1.13$ and $M_G \approx 0.45$. We then excise systems close to this peak (see Fig. \ref{fig:hrd_gaia}). We further remove those with $P < 300$ days, to minimize pollution by less evolved red giants. For the AGB stars that may co-exist with the same radii (especially at the large end) and luminosities as RGBs, we ignore their contribution on account of their relatively short life-span and relatively small number. 

\begin{figure}
    \centering
    \includegraphics[width=0.45\textwidth,trim=5 5 40 30,clip=]{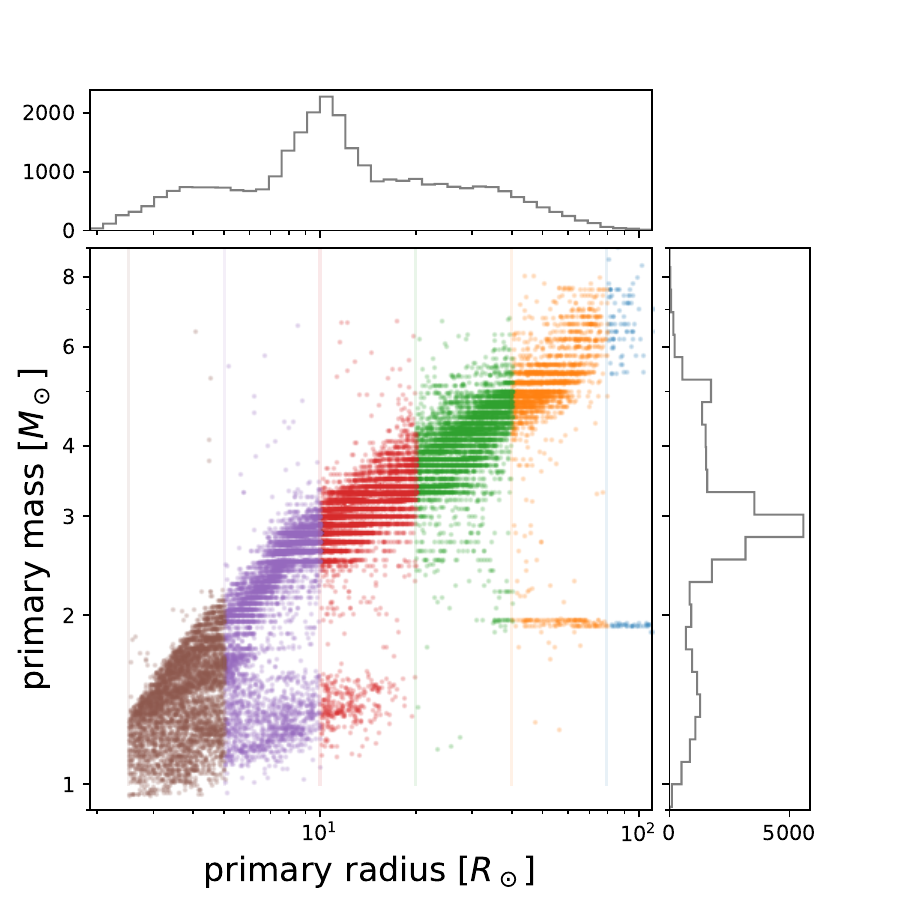}
    \caption{Radii and inferred masses for the primary stars in the giant binaries. There is a strong correlation between radius and mass. Red clump stars cause the large excess around $R/R_\odot \sim 10$. Their mass determination is difficult and may be erroneous.
    }
    \label{fig:mass}
\end{figure}

Gaia DR3 provided mass estimate for the giant primaries. They range from $1 M_\odot$ to $10 M_\odot$ and are shown in Fig. \ref{fig:mass}. These masses are obtained from model fitting and indicate that more luminous giants are also more massive. Among these mass estimates, those for red-clump stars are uncertain and can fail. We will be using these mass estimates later.

\begin{figure*}
    \centering
    \includegraphics[width=\textwidth]
    {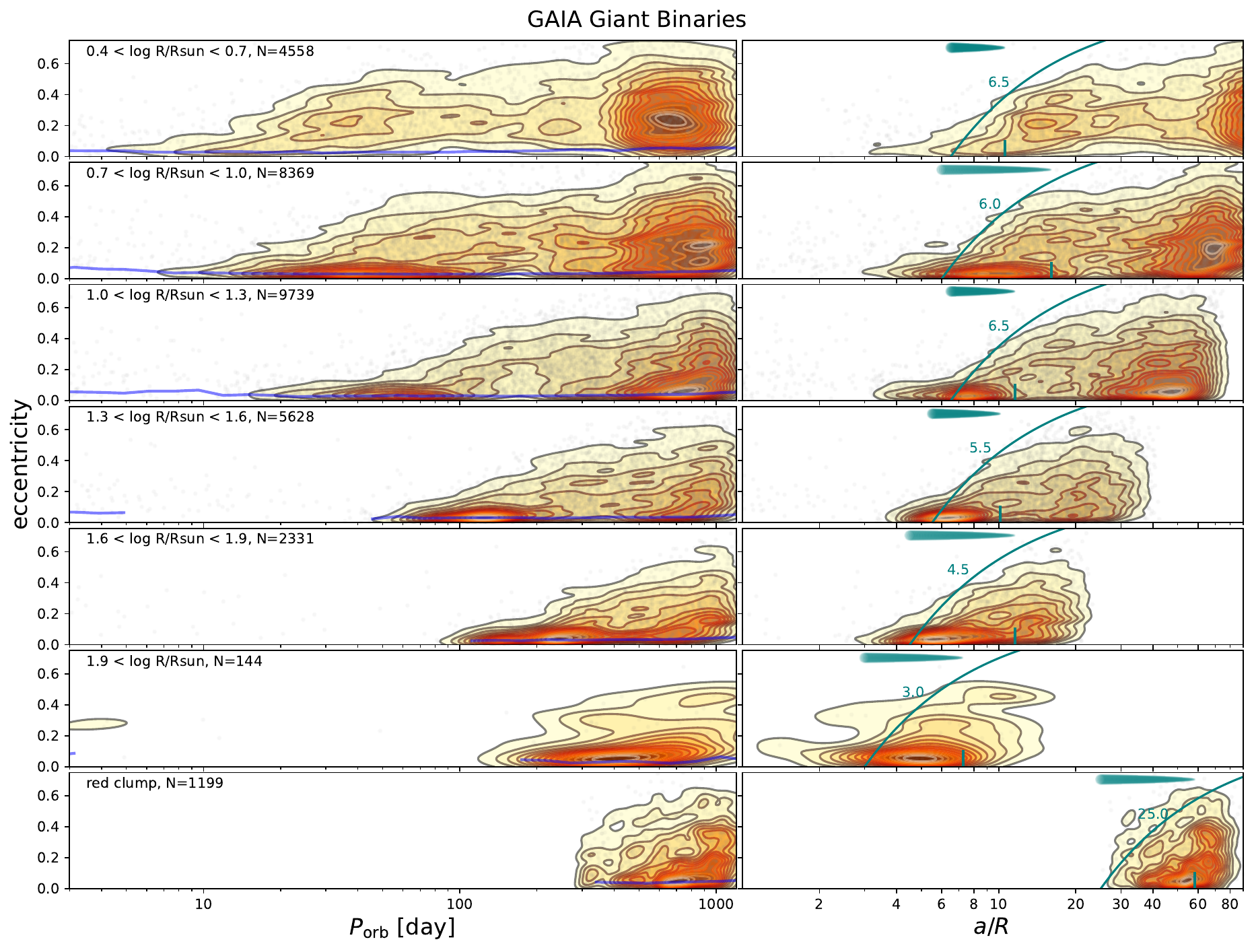}
    \caption{Eccentricity distributions of  Gaia giant binaries, represented by kernel density estimates and separated into groups by the primary radii. In the left column the x-axis shows period, while in the right column we convert period to the ratio $a/R$. The eccentricity distribution is more sharply defined in the latter space. We measure the reach of tidal circularization in this space, following  the same procedure  as in Fig. \ref{fig:mainsequence}. The shrinking arrows indicate both the eccentricity upper envelopes (teal curves, with constant values of $\beta = a(1-e)/R$ indicated by text annotations,  which largely follow the lowest contour at $0.05$) and the edges of the cool islands (vertical bars).
   }
   \label{fig:redgiants}
\end{figure*}

The orbital eccentricities of giant binaries are presented in Fig. \ref{fig:redgiants},  again separated into groups by their primary radii and plotted as a function of binary separation. The measurement uncertainties on eccentricities are typically $2-4\%$,  depending on stellar brightness and period. Here we offer some observations.

Analogous to main-sequence stars, giant binaries also exhibit an upper envelope in the eccentricity-period space. As Fig. \ref{fig:redgiants} shows, this upper envelope moves outwards in period as the primary star ascends the giant branch. The corresponding peri-centre period evolves from $\sim 6$ days on the main-sequence, to $\sim 200$ days in the red-clump stage.  Such an evolution \citep[also apparent in the oscillating subsample identified by ][]{Beck2024} is clearly caused by events on the RGB, with tidal evolution being the most likely culprit. 

This evolution can be better viewed with a different measure of orbital separation, the ratio between semi-major axis and stellar radius ($a/R$). The tidal physics depends sensitively on this ratio. To convert orbital periods to $a$, we adopt primary masses as provided by the  Gaia catalogue (Fig. \ref{fig:mass}), and assume a total system mass of $1.5\times$ the primary mass. The right panels of Fig. \ref{fig:redgiants} show the results of this conversion. In this space, we observe that the tidal features are much more stationary and more sharply defined. Defining an upper envelope in eccentricity by a constant peri-centre distance, $\beta = a(1-e)/R$, we find that $\beta$ evolves from $\sim 6.5$ in the sub-giant stage to $\sim 3$ near the tip of the RGB. In comparison, this value is $\sim 12$ for a $1.5M_\odot$ primary on the main-sequence (a period of 6 days). So while tidal dissipation continues to operate, it is able to reach proportionately smaller distances in more evolved stars.

In Fig. \ref{fig:redgiants}, one also observes the same `cool islands' as in main-sequence binaries,  circular orbits that extend beyond the above $a/R$ threshold. Compared to that on the main-sequence (a cool island extending to $a/R \sim 23$ for a $1.5 M_\odot$ primary), these are less extensive and only reach $a/R \sim 10$ across the RGB. Following the same procedure as for main-sequence stars (Fig. \ref{fig:mainsequence}), we use both the upper-envelopes and cool-island edge to define the reach of tidal circularization. These are shown as shrinking arrows in Fig. \ref{fig:redgiants}, which we will compare against theoretical calculations. Uncertainties on these values can arise from sample variance, uncertain binary masses and measurement errors. They are hard to quantify at this moment.

Lastly, one also observe secondary `cool islands' in more luminous giants, appearing at periods beyond $500$ days. This population is most likely contributed by red clump stars that have descended from the tip of RGB. They were circularized to long periods while the primaries were large. They appear here because they have the same radii as first-ascent RGBs.

\section{Theoretical model}\label{sec:mod}
This section describes the numerical tools that we use to forward model the tidal evolution of red giant binaries. In short, we compute dissipation rates from the linear tidal responses of realistic stellar models. In contrast to main-sequence binaries, the expansion of red giants decreases stellar dynamical frequencies. Consequently tidal forcing can resonate with stellar oscillations modes (f-modes), and stars may be tidally spun up to near break-up. These complications lead us to consider `dynamical' tidal models that include the frequencies of tidal driving, in addition to standard `equilibrium' tidal theories that posit a hydrostatic bulge. 

\subsection{Stellar model}
As a representative system, we consider a binary with a primary star with an (initial) mass $M=1.5M_\odot$, and a secondary with mass $M'=M_\odot.$ We ignore evolution (and tides) in the secondary, and evolve the primary's internal structure using Modules for Experiments in Stellar Astrophysics \citep[MESA; ][]{Paxton2011, Paxton2013, Paxton2015, Paxton2018, Paxton2019, Jermyn2023}. We use MESA version \texttt{r24031}, and only modify the initial mass to $1.5 M_\odot$  in the \texttt{1M\_pre\_ms\_to\_wd} test suite.  The blue curve in Fig. \ref{fig:hrd_gaia} compares the evolution of this model \citep[converted from luminosity and effective temperature to color and magnitude following procedure in the Gaia Flame package,][]{flame2007} to our Gaia sample. 

The \texttt{1M\_pre\_ms\_to\_wd} suite prescribes mass loss on the RGB and AGB as a function of stellar luminosity, radius, and mass \citep{Reimers1975,Bloecker1995}. This mass loss affects the star's rotation and its tidal flows. Tidal spin-up could  in turn invalidate these mass-loss prescriptions, which are mostly calibrated from single-star observations. These interactions should be considered in future.

Our tidal calculations are based on the evolutionary track of a single $1.5 M_\odot$ star. Such a procedure is justified; while the observed red giant binaries have different masses (Fig. \ref{fig:mass}), the radius of a red giant is much more relevant to its tidal evolution than its mass. Moreover, red giants of similar radii should have similar luminosities and similar mean densities. These similarities yield comparable turbulence properties.

\subsection{Governing equations}
Consider a star characterized by background density $\rho_0(r),$ pressure $P_0(r),$ gravitational potential $\Phi_0(r),$ and velocity field ${\bf u}_0=\boldsymbol{\Omega}\times{\bf r}=r\sin\theta\Omega\hat{\boldsymbol{\phi}}$ in spherical polar coordinates $(r,\theta,\phi)$, where $\Omega$ is a uniform rotation rate. The linearized equation of motion governing the fluid response to a small-amplitude tidal driving force ($-\nabla U$) can be written in the co-rotating frame of the star as
\begin{equation}\label{eq:eom1}
    \frac{\partial^2\boldsymbol{\xi}}{\partial t^2}
    +2\boldsymbol{\Omega}\times
    \frac{\partial\boldsymbol{\xi}}{\partial t}
    =\frac{\nabla\rho_0}{\rho_0}
    (c_s^2\eta-h)
    -\nabla(h + \delta\Phi + U)
    +{\bf f}_d.
\end{equation}
Here $\boldsymbol{\xi}$ is the Lagrangian displacement vector, which in a rigid rotator is related to the Eulerian velocity perturbation ${\bf v}$ by ${\bf v}=\partial\boldsymbol{\xi}/\partial t.$ Meanwhile, $c_s^2=\text{d}P_0/\text{d}\rho_0$ gives the equilibrium sound speed, which deviates from the adiabatic sound speed ($c_A^2=\Gamma_1P_0/\rho_0$, where $\Gamma_1$ is the first adiabatic exponent) in any regions with stable (or unstable) entropy stratification. Additionally $\eta=\delta\rho/\rho_0$, $h=\delta P/\rho_0$, and $\delta \Phi$ describe Eulerian perturbations to the density, pressure, and gravitational potential, respectively. Ignoring non-adiabatic heat exchange, these variables satisfy the relations
\begin{equation}
    \rho_0\eta=-\nabla\cdot(\rho_0\boldsymbol{\xi}),
\end{equation}
\begin{equation}\label{eq:en1}
    h=c_A^2\eta 
    +(c_A^2\nabla\ln\rho_0-\boldsymbol{\mathcal{G}})\cdot\boldsymbol{\xi},
\end{equation}
\begin{equation}\label{eq:poi1}
    \nabla^2\delta\Phi 
    =4\pi G\rho_0\eta,
\end{equation}
where $\boldsymbol{\mathcal{G}}=\rho_0^{-1}\nabla P_0$ is the effective gravity of the equilibrium star, and $G$ is the gravitational constant.

In Equation \eqref{eq:eom1}, ${\bf f}_d$ describes  dissipative forces per unit mass. We include a viscous force ${\bf f}_d=\rho_0^{-1}\nabla\cdot(2\rho_0\nu_c{\bf S}),$ where $\nu_c$ is an effective kinematic viscosity due to turbulent convection (see below) and 
\begin{equation}\label{eq:SS}
    {\bf S}=
    \frac{1}{2}\left[
        \nabla{\bf v}
        +(\nabla{\bf v})^T
        -\frac{2}{3}(\nabla\cdot{\bf v}){\bf I}
    \right].
\end{equation}

\subsection{Viscous dissipation}\label{sec:nuc}
Classical tidal theories \citep{Zahn1977} assume that convective turbulence provides an effective viscosity ($\nu_c$) that is determined by the length and timescales of convective eddies, modulated by the ratio of the (rotating frame) tidal frequency ($\omega$) to the convective turnover frequency. We adopt the scaling inferred from local simulations by \citet{Duguid2020}:
\begin{align}\label{eq:numod}
    \nu_c
    =\begin{cases}
        5\ell_cv_c, & |\omega|/\omega_c<10^{-2} \\
        \frac{1}{2}\ell_cv_c
        \left(\omega_c/|\omega|\right)^{1/2}, 
        & 10^{-2}\leq|\omega|/\omega_c\leq 5\\
        \frac{25}{\sqrt{20}}\ell_cv_c 
        \left(\omega_c/|\omega|\right)^2, 
        & 5<|\omega|/\omega_c
    \end{cases}
\end{align}
where $\ell_c$ and $v_c$ are convective mixing length and velocity scales in the evolving star, and $\omega_c=v_c/\ell_c$. For fast tides ($\omega \gg \omega_c$), viscosity is suppressed following  the scaling proposed by \citet{Goldreich1977}. There remains some controversy about this suppression  \citep{Goodman1997,Terquem2021,Terquem2023,Barker2021}, but it is relatively unimportant to our calculations. The tidal periods of concern are much longer on the RGB (hundreds of days) than  on the main-sequence (days), while  convective turn-over times remain comparable (Fig. \ref{fig:nuc}). 

The volume-integrated rate of energy dissipation due to the effective viscosity $\nu_c$ is then
\begin{align}\label{eq:D}
    \mathcal{D}
    &=\int_{V^*}2\rho_0\nu_c{\bf S}:{\bf S}\text{d}V.
\end{align}
The tensor contraction ${\bf S}:{\bf S}$ can be computed from the velocity field ${\bf v}=\partial\boldsymbol{\xi}/\partial t,$ with $\boldsymbol{\xi}$ in turn calculated as described in \S \ref{sec:eqm}.

\subsection{Methods for computing the tidal flow}\label{sec:eqm}
The tidal potential appearing in Equation \eqref{eq:eom1} can be written approximately as
\begin{equation}\label{eq:U0}
    U\simeq\sum_{m=-2}^2
    \sum_{k=-\infty}^\infty
    U_{mk}\left(\frac{r}{R}\right)^2
    Y_{2m}
    \exp[-\text{i}\omega_{mk}t].
\end{equation}
Here $U_{mk}$ are coefficients that depend on the orbital semi-major axis $a$ and eccentricity $e$ (see  \autoref{app:EAM}), $R$ is the (equatorial) radius of the tidally perturbed star, $Y_{2m}$ is a spherical harmonic of degree $2$ and order $m$, and both $m$ and $k$ are integers. The tidal frequency associated with each  component of this Fourier decomposition is $\omega_{mk}=k\Omega_0-m\Omega$, where $\Omega_0=[G(M+M')/a^3]^{1/2}$ is the orbital mean motion. This frequency is the only place where we incorporate the effects of stellar rotation into our tidal calculations. For $e=0$, the only nonzero coefficients $U_{mk}$ are those with $m=k$. Eccentric orbits, on the other hand, involve driving by potentials with many more tidal frequencies. 

We adopt a linear approximation, assuming that the overall tide can be described as the superposition of the (infinitely many) responses $\boldsymbol{\xi}_{mk}$ to individual potentials, $\boldsymbol{\xi}=\sum_{m,k}\boldsymbol{\xi}_{mk}$. Additionally we assume that each linear response takes the same time dependence as its driving tidal potential (so that $\partial\boldsymbol{\xi}_{mk}/\partial t=-\text{i}\omega_{mk}\boldsymbol{\xi}_{mk}$ for each $m,k$). Each response and its time-averaged dissipation rate can  then be computed independently. We omit the indices $m$ and $k$ unless needed. 

We adopt two main simplified models: `equilibrium,' and `f-mode' tides. The equilibrium tidal model, which we review in \autoref{app:eqm}, assumes that the spatial dependence of every linear tidal response is described by the solution to Equation \eqref{eq:eom1} in the limit of vanishing tidal frequencies $\omega\rightarrow0$ \citep[i.e., Equation \ref{eq:eom1} with the lefthand side set to zero; ][]{Zahn1977}. This approximation is reasonable when the tidal frequencies $\omega_{mk}$ are all $ \ll \omega_{\rm dyn}=[GM/R^3]^{1/2}$, and when they do not resonate with any natural oscillations of the star.

In red giants, tidal frequencies of concern are only somewhat  larger than,  or even comparable to  the stellar dynamical frequencies. So the approximation of equilibrium tides is  questionable. For our default calculation, we adopt a model that accounts for finite tidal frequencies, or, the so-called `dynamical tides'. Appendix \ref{app:dyn} describes our approach, which is similar to that adopted by \citet{Vick2020}. In short, we approximate the tidal response as a truncated expansion in normal mode oscillations of the  evolving primary star, without insisting $\omega \rightarrow 0$. This approach accounts for both resonant and non-resonant  driving of the stars' fundamental and acoustic modes. We refer to this model as `f-mode' tides, to differentiate it from other dynamical tide calculations focused  on internal gravity \citep[e.g.,][]{Zahn1977} and inertial \citep[e.g.,][]{Ogilvie2013} waves. 

\subsection{Secular orbital evolution} 
Under tidal dissipation, the orbital semi-major axis $a$ and eccentricity $e$ evolve as (see Appendix \ref{app:EAM}) \begin{align}\label{eq:adot}
    \frac{\dot{a}}{a}
    &=-\frac{\Omega_o\epsilon_t}{2}
    \sum_{k=-\infty}^\infty 
    k
    \left(
        |X_{0k}|^2\kappa_{0k}
        +3|X_{2k}|^2\kappa_{2k}
    \right),
\\\label{eq:edot}
    \frac{\dot{e}}{e}
    &=\frac{(1-e^2)}{e^2}\frac{\Omega_o\epsilon_t}{4}
    \sum_{k=-\infty}^\infty 
    \Bigg[
    \\&\hspace{1em}\notag
        3\left(
            \frac{2}{(1-e^2)^{1/2}} - k
        \right)|X_{2k}|^2\kappa_{2k}
        -k|X_{0k}|^2\kappa_{0k}
    \Bigg],
\end{align}
where $\epsilon_t=(M'/M)(R/a)^5$. Meanwhile $X_{m k}=X_{mk}(e)$ are Hansen coefficients that depend on the eccentricity $e$ (as well as the integers $m$ and $k$ appearing in Equation \ref{eq:U0}). Lastly, 
\begin{equation}\label{eq:Dtokap}
    \kappa_{mk}
    =\frac{8\pi G}{5R}
    \frac{\mathcal{D}_{mk}}{\omega_{mk}},
\end{equation}
where $\mathcal{D}_{mk}$ are the (amplitude-normalized and time-averaged) dissipation rates associated with each linear response $\boldsymbol{\xi}_{mk}$. Note that $\text{sgn}[\kappa_{mk}]=\text{sgn}[\omega_{mk}]$ for positive definite dissipation \citep{Ogilvie2013}.

Tidal dissipation changes the primary star's spin angular momentum as
\begin{equation}
\label{eq:Jdot}
    \dot{J}
    =-3E_o\epsilon_t
    \sum_{k=-\infty}^\infty |X_{2k}|^2\kappa_{2k},
\end{equation}
where $E_o=-GMM'/(2a)$ is the orbital energy. We make the simplifying assumption that the star rotates uniformly with rotation rate $\Omega$ and moment of inertia $I$. The dimensionless $I$ is 
\begin{equation}
I =\frac{1}{MR^2} \int_{V^*}\rho_0r^2\sin^2\theta\text{d}V\, .
\end{equation}
We incorporate changes in $I$ associated with stellar evolution (as computed by MESA), but ignore its changes due to rotational evolution. 

Equations \eqref{eq:adot}, \eqref{eq:edot}, and \eqref{eq:Jdot} take a slightly different form than is often presented in the binary literature. However, they reduce to Equations (2), (6), and (9) in \citet{Leconte2010} under the assumption that the imaginary parts of Love numbers can be related to a constant time lag $\tau$ via $\kappa_{mk}=k_2\tau(k\Omega_o-m\Omega),$ where $k_2$ is the real part of a hydrostatic Love number of degree $2$. Our equations provide a more flexible framework for incorporating an arbitrary dependence of the coefficients $\kappa_{mk}$ on the tidal frequencies $\omega_{mk}.$ This is necessary to capture the frequency dependence of an eddy viscosity, or the resonant driving of stellar oscillation modes. 

Equations \eqref{eq:adot}-\eqref{eq:Jdot} presuppose that $M$ and $M'$ remain unchanged, while in reality our primary star loses mass through winds as it evolves through the red giant and asymptotic giant phases. We assume that all such mass is lost from the system, and approximate its effect on semi-major axis and stellar angular momentum through two additional terms, $\dot{a}_W/a=-\dot{M}/(M + M'),$ $\dot{J}_W=(2/3)\dot{M}\Omega R^2,$  where negative $\dot{M}<0$ describes mass loss \citep{Hurley2002}. This simplified treatment should reasonably approximate the order of magnitude of these effects.

The theoretical calculations of this paper center on numerical integrations of Equations \eqref{eq:adot}-\eqref{eq:Jdot}. We use the \texttt{BDF} integrator provided in \texttt{scipy.integrate}, and adaptively limit the maximum timestep to be no larger than that in MESA. At each timestep, we compute dissipative coefficients $\kappa_{mk}$ for all values of $m$ and $k$ provided $|X_{mk}(e)|>10^{-10}.$ 

\section{Results}\label{sec:res}
\subsection{Single star evolution}\label{sec:sstar}

We first review stellar properties that are of relevance to tidal dissipation.

\begin{figure}
    \centering
    \includegraphics[width=\columnwidth]{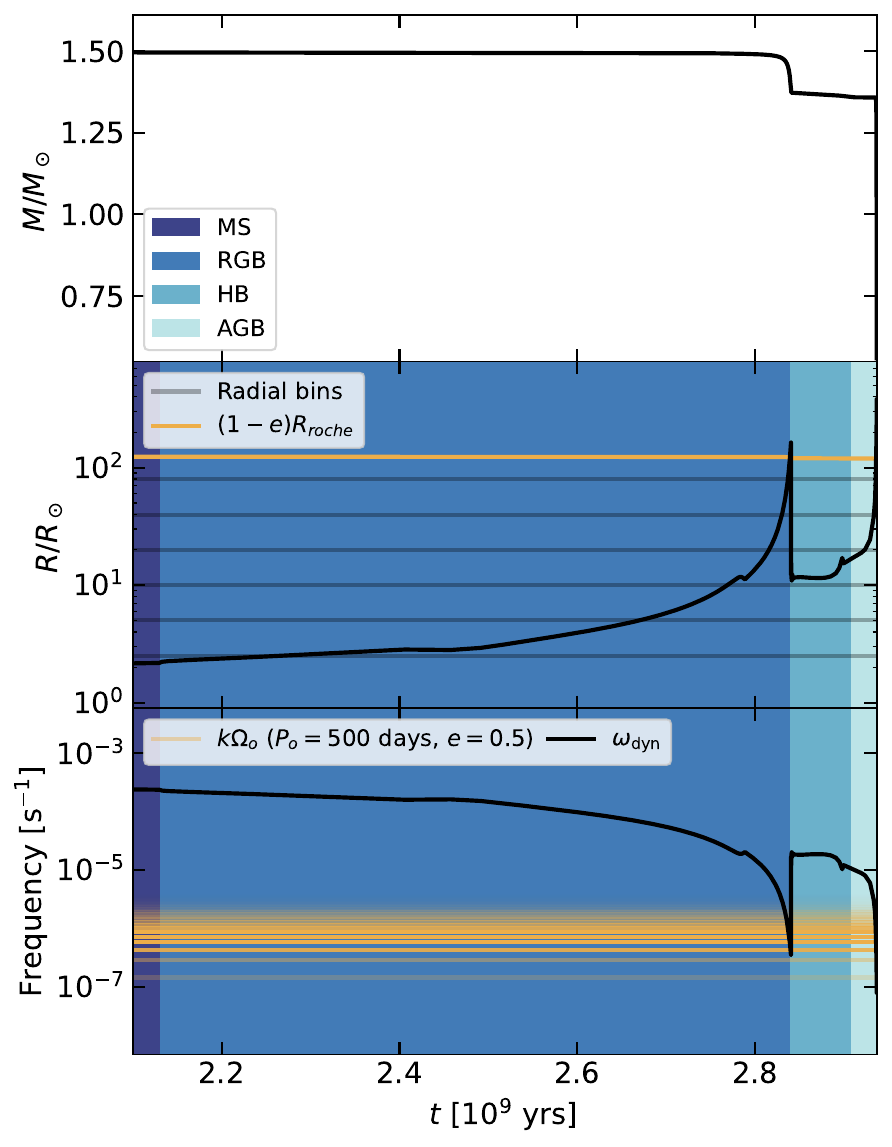}
    \caption{Evolution of mass (top), radius (middle), and relevant frequencies (bottom) for a $1.5M_\odot$ star. The background colors indicate different evolutionary stages. Rapid radial expansions, accompanied by mass-loss, occur near the end of RGB and AGB (grey lines in the middle panel indicate the radial bins applied to our sample). 
    Horizontal yellow lines indicate the Roche radius (middle panel) and tidal driving frequencies (lower panel, opacities scaled by tidal driving amplitudes) for a $500$d, $e=0.5$ binary orbit (with a secondary mass of $1.0M_\odot$). 
    }
    \label{fig:RM}
\end{figure}

Fig. \ref{fig:RM} plots the mass, radius, and relevant frequencies for our $1.5M_\odot$ primary star as it evolves past the main sequence toward its final white dwarf stage. We focus on tidal evolution during the RGB and HB (red clump) phases. 

Compared to its main-sequence progenitor, a giant star experiences stronger tidal driving first of all because the factor $\epsilon_T$ in Equations \eqref{eq:adot}-\eqref{eq:Jdot} scales as $(R/a)^5$. The tidal driving also strengthens as the dynamical frequency ($\omega_{\rm dyn}=[GM/R^3]^{1/2}$) of the star decreases and approaches the tidal frequencies (given by $k \Omega_0$ in the non-rotating frame; see the bottom panel of Fig. \ref{fig:RM}). So for systems with substantial eccentricities, the dynamical frequency can resonate with the dominant ($k>2$) tidal frequencies as the primary approaches the tip of the RGB. In such systems, an equilibrium tide model is less accurate.

\begin{figure}
    \centering
    \includegraphics[width=\columnwidth]{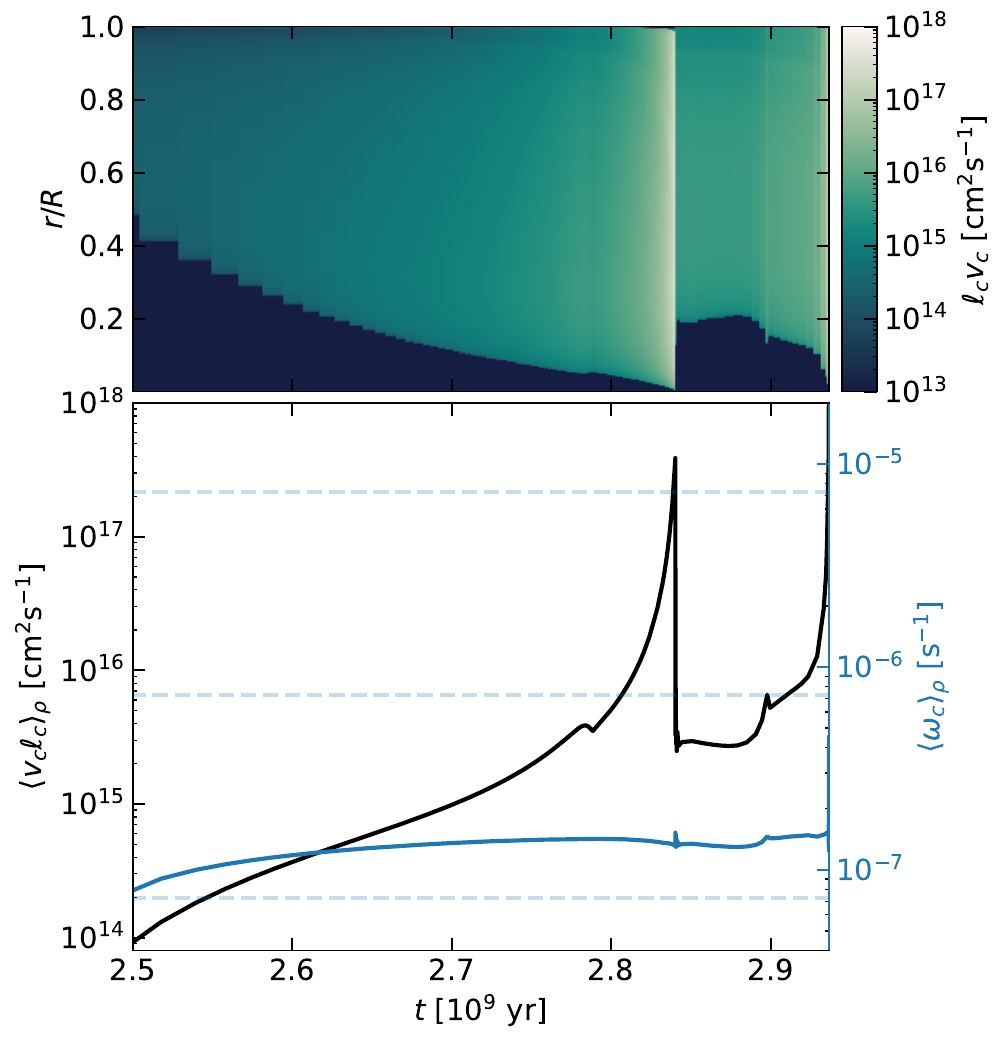}
    \caption{Properties of turbulent convection from the middle of the RGB to end of the AGB. The top colormap shows the product of mixing length and convective velocity ($\nu_c\propto\ell_cv_c$; see Equation \ref{eq:numod}) as a function of both time and fractional radius. The dark region is radiative. The bottom panel plots the mass-weighted averages $\langle\ell_cv_c\rangle_\rho=\int_V\rho_0\ell_cv_c\text{d}V/\int_V\rho_0\text{d}V$ (black) and $\langle \omega_c\rangle_\rho$ (blue), where $\omega_c = v_c/\ell_c$. Faint dashed lines indicate frequencies corresponding to $10,100,$ and $1000$ day periods. 
    Near the tips of RGB and AGB, mixing length theory predicts enhanced turbulent viscosity that contributes to intense dissipation during these stages.
    }
    \label{fig:nuc}
\end{figure}

Another important distinction of giant stars is their turbulent state. A larger luminosity and lower mean density bring about faster convective velocities $v_c$, as the star expands and increases in scale height ($l_c$). In combination, we find that the convective frequency (with a mass-weighted average plotted in the lower panel of Fig. \ref{fig:nuc}) remains largely constant throughout the star's evolution, and is comparable to the orbital mean motion of a $\sim 500$ day orbit. This means that for most orbits of concern ($100-1000$days), turbulent viscosity does not suffer strong suppression due to `fast tides.' At the same time, we observe a dramatic rise in eddy viscosity during the RGB (and AGB), as is shown in Fig. \ref{fig:nuc}. This rise is even more significant in the late stages, and  leads to strong dissipation during those short intervals.

Lastly, the expansion of red giants introduces a new complication, Roche-Lobe-Overflow (RLOF). The expected Roche radius for a circular orbit ($R_{\rm roche}$) is given approximately by \citep{Eggleton1983}
\begin{equation}\label{eq:roche}
    R_\textrm{roche}
    \simeq\frac{0.49a(M/M')^{2/3}}{0.6(M/M')^{2/3} + \ln[1 + (M/M')^{1/3}]}.
\end{equation}
We crudely take $R\geq(1-e) R_{\rm roche}$ as the criterion for RLOF in an  eccentric orbit. As Fig. \ref{fig:RM} indicates, most eccentric binaries with orbital periods $\lesssim 1000$d will encounter RLOF, independent of any tidal theory. We make no attempt to model RLOF  or common-envelope (CE) evolution in this paper. We instead focus on the efficiency of tidal eccentricity damping prior to their onset. We can then predict the orbital eccentricity at the onset of RLOF. Previous treatments \citep[e.g.,][]{Hurley2002}, using constant time lag and equilibrium tides, predict that binaries should be circularized before the primary star overflows its Roche lobe. However, the validity of equilibrium tides is questionable for red giants (see above).
And recent calculations of f-mode tides in massive stars \citep{Vick2021} find that many systems may enter RLOF with significant residual eccentricity. 
\begin{figure}
    \centering
    \includegraphics[width=\columnwidth]{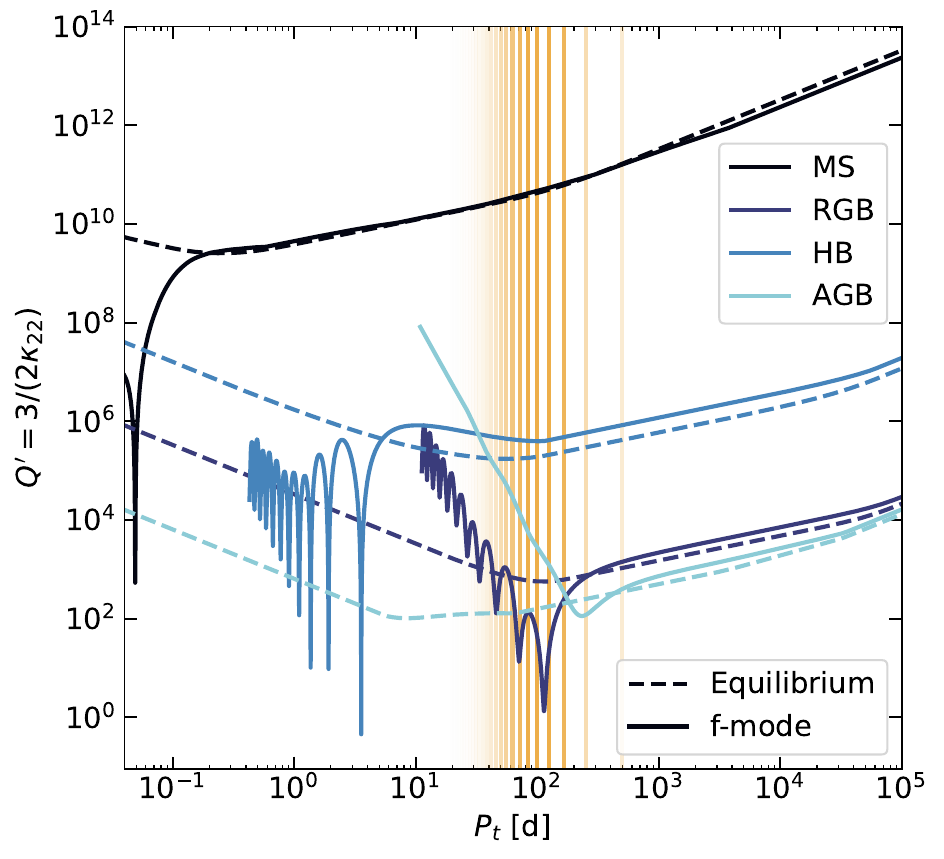}
    \caption{Tidal quality factors (inversely related to the dissipative coefficients appearing in Equations \ref{eq:adot}-\ref{eq:Jdot}), computed as functions of tidal period ($P_t = 2\pi/\omega$), at the end of each evolutionary stage. Dashed lines indicate `equilibrium' tidal calculations, while solid lines show `f-mode' tidal calculations. The yellow lines indicate relevant tidal periods (and corresponding driving amplitudes), as in the bottom panel of Fig. \ref{fig:RM}. The f-mode and equilibrium calculations disagree when the tidal frequency becomes comparable to the stellar dynamical frequency.}\label{fig:Imk22}
\end{figure}

\subsection{Tidal dissipation}
We use the above stellar models to compute tidal dissipation rates as the star ages. In the following, we will present results using the f-mode tides. Results obtained with equilibrium tides (shown in Fig. \ref{fig:evpsnap} of Appendix \ref{app:samp}) are largely similar.
Fig. \ref{fig:Imk22} shows four snapshots of this dissipation, taken at the end of each major evolutionary phase (from main-sequence to AGB). We plot the commonly used `effective tidal quality factor' \citep[inversely proportional to the $m=k=2$ dissipative coefficient appearing in Equations \ref{eq:adot}-\ref{eq:Jdot}; e.g., ][]{Ogilvie2014} as a function of orbital period, for both equilibrium and f-mode tides. For f-mode tides, we terminate the curves at the shortest tidal periods allowed by our mode expansion (e.g., $P_t\simeq10$ days at the end of the RGB). 

As expected, we find the strongest tidal dissipation (i.e., smallest values of $Q'$) during  periods of largest radial expansion (tips of the RGB and AGB). These enhancements are due to changes in both the star's structure and to its eddy viscosity. They lead to significant tidal evolution during the fast expansion phases of the RGB and the AGB, despite their brevity.

Fig. \ref{fig:Imk22} also highlights the differences between the f-mode and equilibrium tidal models. 
Results from f-mode tidal calculations asymptote toward equilibrium predictions at long periods, as is expected. But at short periods they contain more structure due to resonances with fundamental and acoustic modes, and can differ from the equilibrium results by orders of magnitude. At short periods (high $\omega$), f-mode tidal dissipation weakens more dramatically with increasing $\omega$ than equilibrium tidal dissipation (which weakens as $P_t^2$ due to the high-frequency suppression of turbulent viscosity). This steeper drop-off is due to the decreasing Lorentzian wing at frequencies larger than the f-mode's natural oscillation frequency. Because equilibrium tidal models assume $\omega=0$ when computing the tidal flow (i.e., the displacement vector $\boldsymbol{\xi}$), they cannot capture this `dynamical' high-frequency suppression. 

\subsection{Initial binary orbits}\label{sec:ics}
Since the giant binaries evolve from main-sequence binaries, the latter provide the initial orbital distributions for our evolution. 

The period distribution of solar-type binaries is commonly reported to be log-normal with a maximum at $P\sim 10^5$ days \citep[see, e.g.][]{Moe2017}. We draw initial values of ${\rm log}_{10}P$ from a normal distribution with a mean of $5$ and a standard deviation of $2.3.$

For the eccentricity distribution, we recently reported that Gaia main-sequence binaries obey a Rayleigh distribution in eccentricities \citep{Wu2024}: 
\begin{equation}
\frac{dN}{de} \propto e \,\, 
{\exp[- e^2/(2\sigma_e^2)]\, ,}
\label{eq:rayleigh}
\end{equation}
with a Rayleigh mode $\sigma_e \approx 0.30$ that appears to be largely independent of stellar spectral type, and applies to binaries with periods from $\sim 20$ to $1200$ days. Binaries inward of $20$days appear affected by tides (Fig. \ref{fig:mainsequence}) and have smaller eccentricities. As we are only concerned with giant binaries which have longer periods, we can safely adopt the above Rayleigh distribution.

\begin{figure}
    \centering
    \includegraphics[width=\columnwidth]{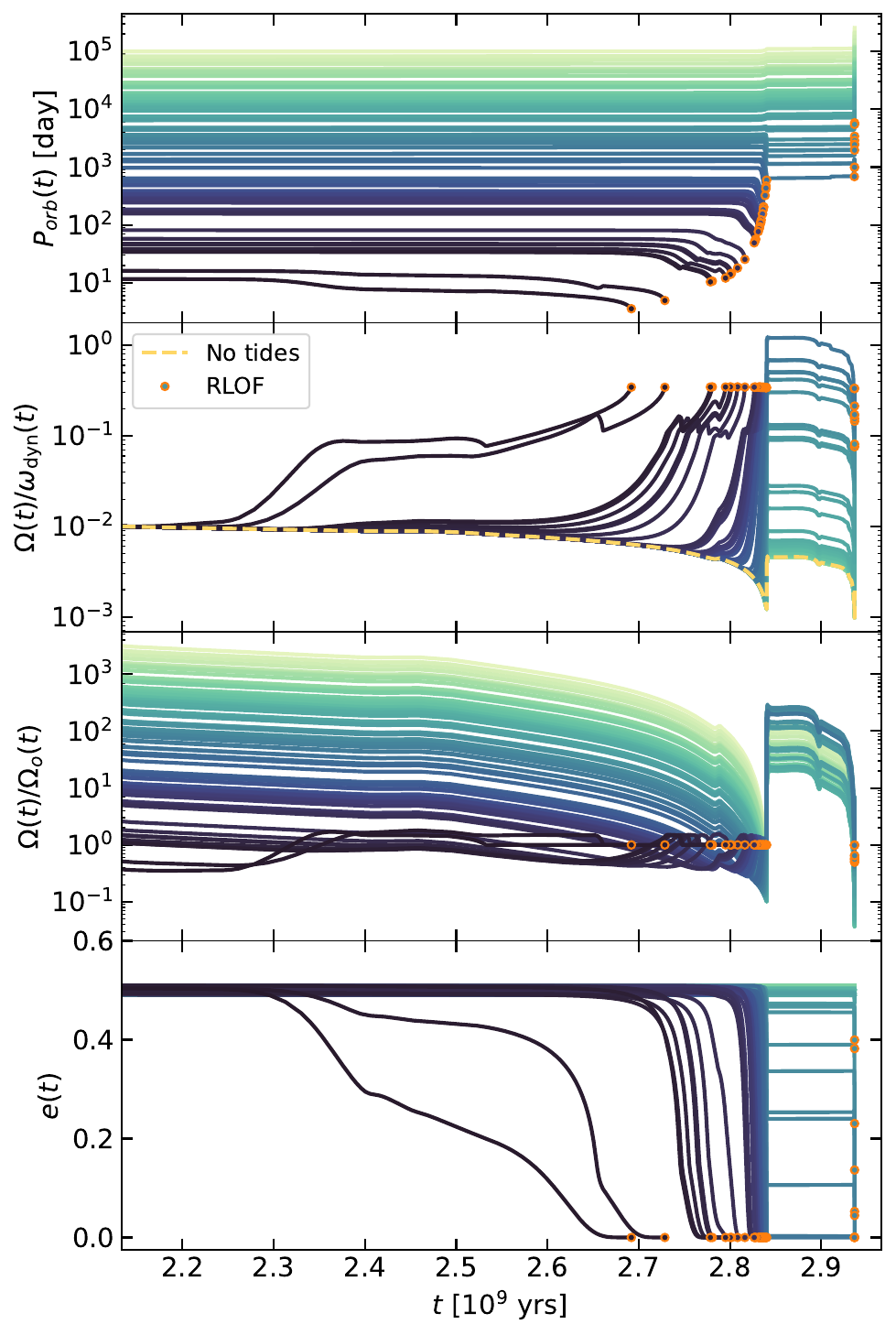}
    \caption{Orbital evolution driven by f-mode tides. From top to bottom, the panels show orbital period, stellar rotation rate (relative to both the dynamical frequency and the orbital mean motion), and eccentricity for some representative initial conditions (light/dark line colors indicate long/short orbital periods). Orange circles indicate RLOF. Eccentricity damping takes place almost exclusively during radial expansion on the RGB and AGB. The spin evolution of a single star (yellow dashed line) matches spin evolution in the widest binaries (light-green color).
    }
    \label{fig:e5_ts}
\end{figure}

\begin{figure*}
    \centering
    \includegraphics[width=\columnwidth]{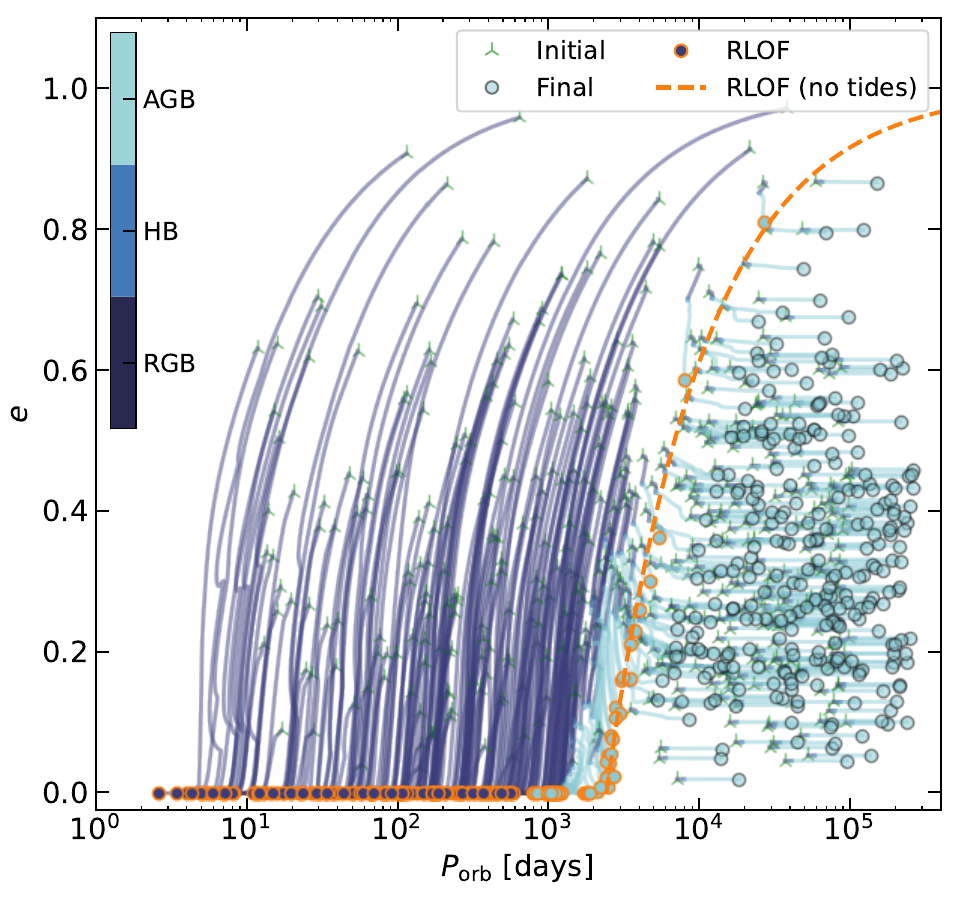}
    \includegraphics[width=\columnwidth]{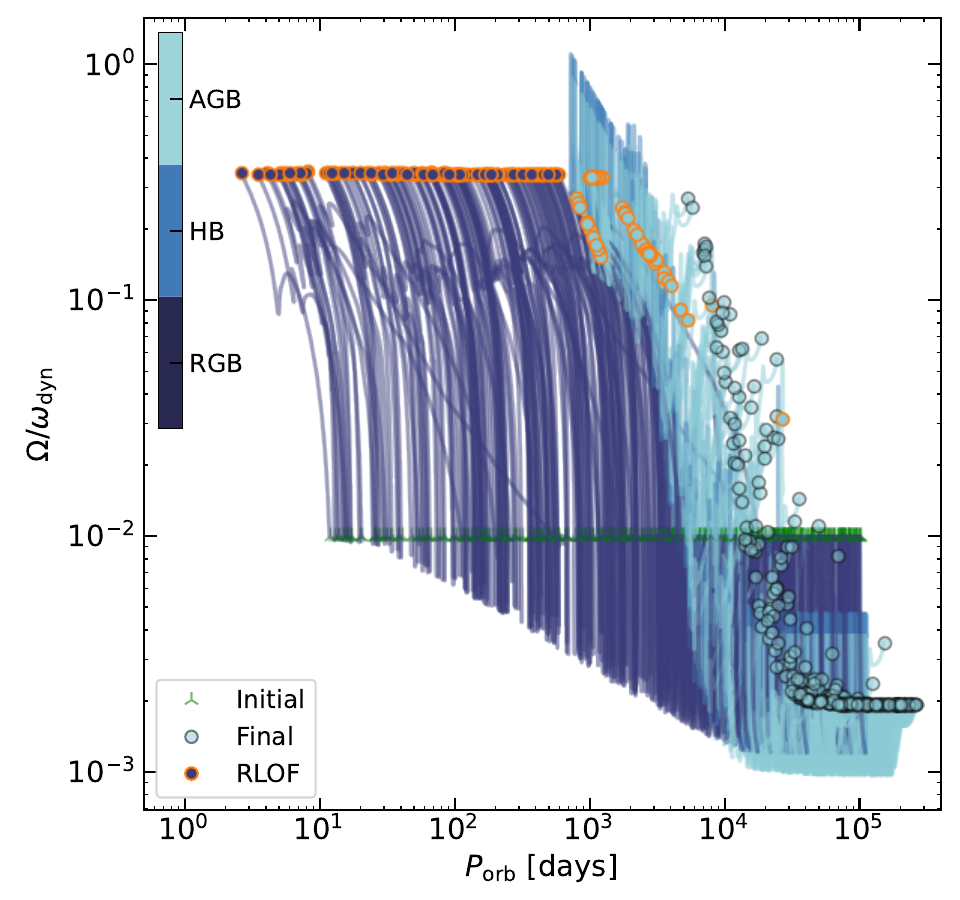}
    \caption{Evolution of orbital eccentricity (left panel) and primary spin (right panel, normalized by $\omega_{\rm dyn}$), as functions of orbital period, for an assortment of initial conditions (green tri-stars) and integrated with f-mode tides. Line colors indicate evolutionary stages, and orange outlined points indicate end points concluding in RLOF. The orange dashed line delineates the boundary of RLOF, in the absence of tides. After the RGB and the AGB, tidal circularization extends to binaries of periods $\sim 1000$ days and $\sim 2000$ days (respectively). In close binaries the primary star's rotation synchronizes with the orbital mean motion (approaching $\Omega\sim0.35\omega_{\rm dyn}$ just prior to RLOF). Tides in binaries with $P_\text{orb}$ between $5000$ and $10^4$ days are likely to be strongly affected by spin-up to large fractions of $\Omega/\omega_{\rm dyn}$ during the RGB and AGB (respectively). At larger separations the spin evolution is dominated by mass loss.}\label{fig:param_Pve}
\end{figure*}

\subsection{Orbital evolution}\label{sec:orb}
This section describes the impacts of tidal dissipation on the orbits of giant binaries.

We incorporate on-the-fly calculations of dissipation coefficients like those shown in Fig. \ref{fig:Imk22} into the numerical integration of Equations \eqref{eq:adot}-\eqref{eq:Jdot}, interpolating from oscillation mode properties pre-computed for each MESA snapshot (see Appendix \ref{app:dyn}). We assume that our $1.5M_\odot$ primary star is in an orbit with a $1M_\odot$ secondary. We draw initial eccentricities and periods from the Rayleigh and log-normal distributions described above. We additionally adopt an initial rotation rate of $\Omega=0.01\omega_{\rm dyn}$ in all cases, noting that this condition is quickly erased by spin evolution due to tides and mass loss.

We start each integration from the beginning of the RGB, and continue until the end of the AGB, or until the stellar radius surpasses the Roche radius $(1-e)R_\text{roche}$ (with $R_\text{roche}$ given by Equation \ref{eq:roche}). 

Fig. \ref{fig:e5_ts} shows representative evolution due to f-mode tides. Most of the tidal evolution (circularization, shrinkage, synchronization) indeed occurs during the major expansion phases. Orbital shrinkage leads to RLOF for close-in systems. For an orbit with an initial eccentricity of $\sim0.5$, RLOF occurs before the end of the RGB for initial orbital periods $P_\text{orb}\lesssim10^3$ days, and before the end of the AGB for $P_\text{orb}\lesssim3\times 10^3$ days. Many of the binaries with orbital periods between $10^3$ days and $3\times10^3$ days begin RLOF while still significantly eccentric. 

Fig. \ref{fig:param_Pve} shows evolution in eccentricity and spin, plotted vs. orbital period, again for f-mode tides. The left panel also shows the boundary for RLOF in the absence of any orbital evolution. F-mode tidal dissipation has nearly an ``all-or-nothing'' impact on the orbital eccentricity: systems inward of the RLOF contour are circularized before beginning mass transfer, while 
those that avoid RLOF experience minimal circularization, and are instead widened by mass-loss. Binaries starting just above the RLOF boundary provide an exception; they may enter RLOF and begin mass transfer with significant residual eccentricity. The outer boundary for circularization extends to $P_{\rm orb} \sim 1000$ days after the RGB, and to $\sim 2000$ days after the AGB. 

Fig. \ref{fig:param_Pve} also depicts the evolution of the primary star's spin rate. Systems that are close enough to circularize are universally spun up by tides to rotation rates that are comparable with the mean motion, orders of magnitude faster than the rotation rates of single stars. For our binary with a mass ratio of $q=M'/M=2/3$, we obtain $\Omega/\omega_{\rm dyn} \approx 0.3$ at the moment of RLOF.\footnote{
This arises because 
\begin{equation} 
    \Omega/\omega_{\rm dyn}
    \simeq(1 + q)^{1/2}\left[ 
        \frac{0.49q^{-2/3}}
        {0.6q^{-2/3} + \ln(1 + q^{-1/3})}
    \right]^{3/2}
\end{equation}
for a spin-synchronized, circular binary entering RLOF, which evaluates to $\sim0.34$ for $q=2/3$.} Binaries with orbital periods up to $\sim5000$ days experience significant spin-up on the RGB, while those with periods up to $\sim10^4$ days spin-up on the AGB.  

Such rapid rotation has interesting implications. It should strongly impact stellar winds, since an envelope that is partially supported by rotation will be more weakly bound, compared to that of a slowly rotating star. This casts doubt on the validity of our mass loss prescription. Such rapid rotation will also significantly modify the tidal flow and affect the tidal evolution. This deserves more careful study. 

\begin{figure*}
    \centering
    \includegraphics[width=0.925\textwidth]{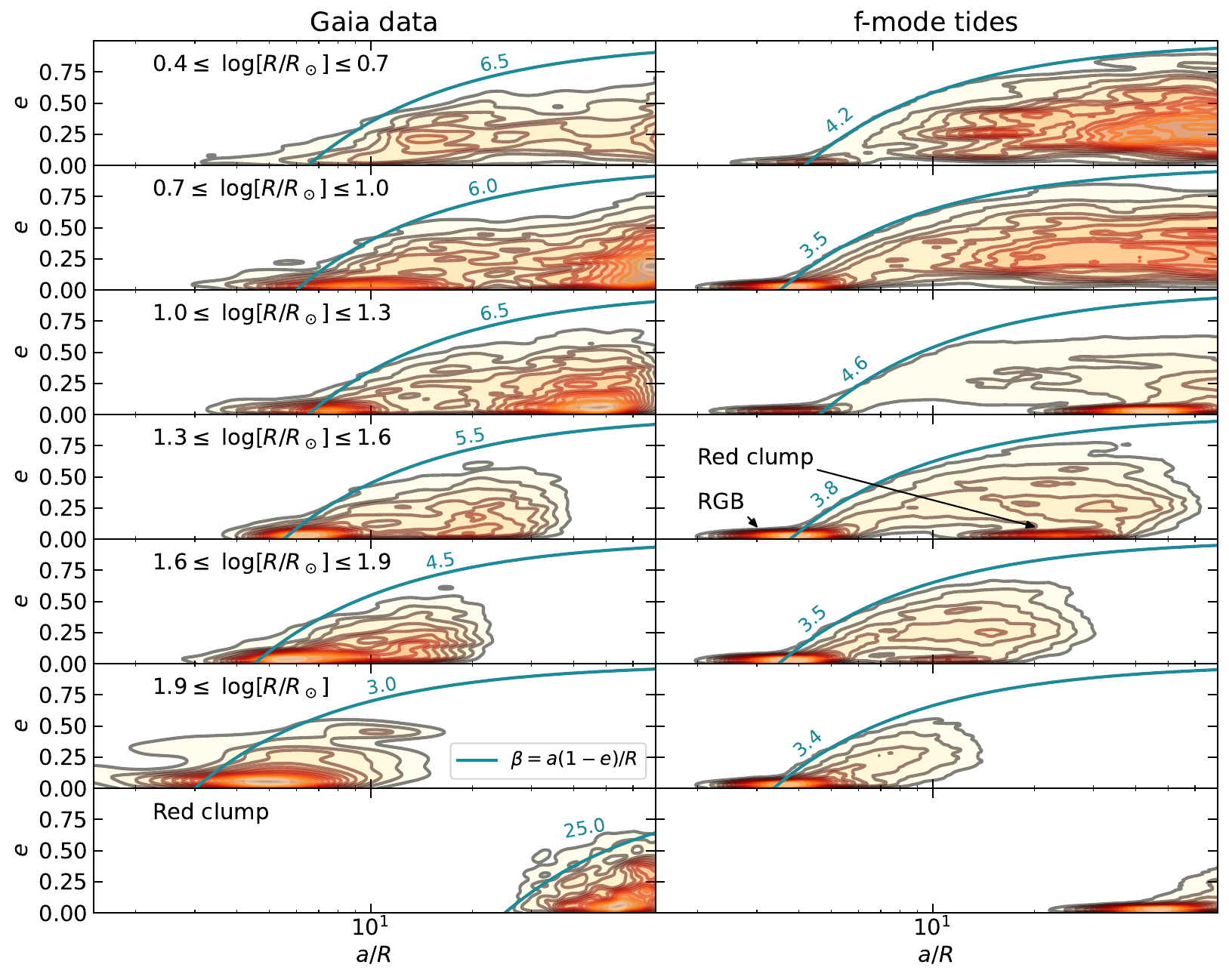}
    \caption{
    Kernel density estimates computed from eccentricity and distance data for Gaia giant binaries (left) and theoretical tidal calculations (right). The panels are split vertically by the radius of the primary star (our proxy for age). The bottom row  singles out red clump stars, although they are also present in other rows as the second ``cool island'' at larger separations. Overall, the theoretical calculations capture qualitative features of the observed eccentricity-distance distributions, but with quantitative disagreement during early stages of the RGB.}\label{fig:fmode_obs_cpr}
\end{figure*}

\subsection{Population-level Results}

Employing all of the elements presented above (\S \ref{sec:sstar}-\ref{sec:orb}), we now construct theoretical samples of binaries at different evolutionary stages. Appendix \ref{app:samp} provides  more details about our sample construction.

Fig. \ref{fig:fmode_obs_cpr} compares  our theoretical eccentricity distributions against Gaia observations. We again split the theoretical sample by primary radii, as  for the Gaia binaries (Fig. \ref{fig:redgiants}), and use $a/R$---rather than orbital period---as our preferred ordinate. 

As the primary stars age and expand in radius, binaries with longer and longer orbital periods can be circularized by tides. (see Fig. \ref{fig:evpsnap}). But the distributions of theoretically predicted binaries remain nearly fixed in $e$ vs. $a/R$ space. Fig. \ref{fig:fmode_obs_cpr} shows that, regardless of primary radii, the upper envelope of eccentricity almost always remains at $\beta \equiv a(1-e)/R \approx 4$. At the same time, a cool island (of circular orbits) extends slightly beyond this, to $a/R \sim 6$. Such a simple description attests to the power of using $a/R$ as the ordinate. We also observe a second cool island at much larger distances. This second island is due to contamination by giants that have descended from the tip of their RGB (red-clump stars).

Calculations using equilibrium tides produce largely similar results. This similarity occurs despite the inclusion of (f- and p-mode) resonances in f-mode tides (see Fig. \ref{fig:Imk22}). These resonances do enhance tidal dissipation strongly when they occur, but close examination reveals that they typically take place too late in the RGB and AGB stages to play a significant role.

\section{Comparison and Discussions}\label{sec:obs}
\begin{figure}
    \centering
    \includegraphics[width=\columnwidth]
    {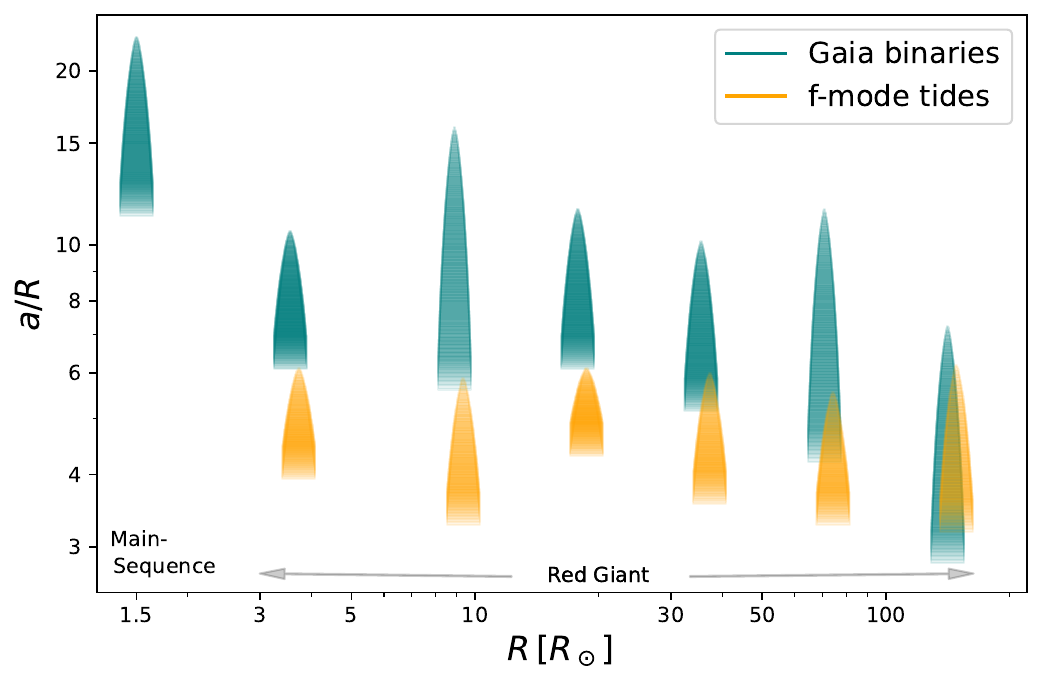}
    \caption{Extents of tidal circularization, obtained from Gaia binaries (teal color, from Fig. \ref{fig:redgiants}), and from our f-mode calculations (orange color, from Fig. \ref{fig:fmode_obs_cpr}), expressed in units of  $a/R$ and plotted as functions of stellar radii (a proxy for evolutionary stage). The extent of each shrinking arrow is defined in Fig. \ref{fig:mainsequence}. While late RGB stars are well described by tidal theory, early RGB stars appear to experience stronger dissipation than predicted. These latter stars also exhibit more extended cool islands than expected, reminiscent of the main-sequence phase.}\label{fig:summary}
\end{figure}

The Gaia giant binaries are already showcased in Fig. \ref{fig:redgiants}. Here we directly compare theory and data (Fig. \ref{fig:fmode_obs_cpr}), using f-mode tides as the default model. Fig. \ref{fig:summary} summarizes our results in terms of critical $a/R$ ratios. While f-mode tides consistently produce eccentricity upper envelopes with $\beta \equiv a(1-e)/R \simeq 4$, the envelopes in the Gaia data evolve from $\beta \approx 6$ in the early RGB to $\beta \approx 3$ near the RGB tip. In addition, the data show cool islands of circular orbits extending to $a/R$ beyond $10$ in the early RGB phase, and to $\sim 6$ in the late phases. In contrast, f-mode tides produce cool islands that always extend to $a/R\approx 6$. This comparison leads us to conclude that an additional process is required to explain circularization in the early RGB, while f-mode tides largely suffice at late stages.

What could be the additional physics at early times?

One possibility is enhanced viscosity. It is possible that we have adopted a prescription for turbulent viscosity that is too conservative. 
During the RGB, the viscous prescription of \citet{Duguid2020} does cause some mild suppression due to `fast tides,' in particular in zones deep below the stellar surface. And this suppression is stronger in the early RGB, as the relevant binary orbits during this period are tighter (Fig. \ref{fig:redgiants}). It is unclear if a different viscous prescription would  help  reconcile the early RGB discrepancy.

The second possibility is internal gravity waves \citep{Zahn1977}. As Fig. \ref{fig:nuc} shows, the inner radiative cores in early RGB stars are more substantial in size. These cores are also highly stratified. Gravity waves that are tidally excited and propagate inward can break in these cores \citep{Goodman1998,Weinberg2017,Guo2023,Esseldeurs2024}. This introduces another source of dissipation that is unaccounted for in our work.

The third possibility is mode locking \citep{Witte1999,Fuller2017}. In the case of pre-main-sequence stars, locking of internal gravity waves has been shown to extend the reach of tidal dissipation \citep{Zanazzi2021}. This process may also operate in the early RGB \citep[see, e.g.,][]{Bryan2024}.

Two other ingredients, not considered here, may also affect  the tides of early RGB binaries. One is  related to rotation.  Rotation introduces a new class of waves called `inertial waves' in stars, and modifies internal gravity modes. The dissipation of inertial waves \citep{Ogilvie2007,Ogilvie2014,Barker2022}
and/or gravito-inertial modes \citep{Xu2017} has been argued to play an important role in some cases. Inertial waves may enhance tidal dissipation in the early RGB, but not in the late RGB, because while both have extended convective envelopes, convective turbulence in the early stage is weaker and may not obliterate inertial-wave resonances \citep[but see][who predict that inertial waves should have a limited impact on the RGB due to a reduced radiative core size]{Beck2018}. The development of differential rotation (here we assume the primary star rotates rigidly) may also affect tidal eccentricity damping in the early RGB.  

The second ingredient is magnetic fields. Stellar magnetic fields can affect internal oscillations \citep{Lin2018,Duguid2024} by absorbing tidal energy into Alfv\'en waves. This can enhance dissipation.

Lastly, there may be extra-tidal effects. For instance, RGB winds may be captured by a binary companion to form a circum-binary disk. This disk may modify the binary eccentricity. There are two arguments against this proposal. First, \citet{Rafikov2016} pointed out that such disks are likely too low in mass to matter; second, such a disk, if it exists, is likely more prominent in late RGB phases (when winds are stronger) than early phases. This may not help resolve the discrepancy we find here.

\section{Conclusions} \label{sec:concl}
Gaia DR3 \citep{Arenou2023} offers us an unprecedented opportunity to scrutinize tidal dissipation as stars evolve along the giant branch. We have made a direct comparison between observed giant binaries and theoretical predictions, the latter obtained through a first-principled calculation of  `f-mode' tides, a non-zero frequency analogue to the well-known equilibrium tidal model. Fig. \ref{fig:summary} summarizes our main results. 

Overall, we find that f-mode tides, together with our adopted eddy viscosity, can circularize binaries out to a fixed ratio of $a/R \approx 4$, regardless of the evolutionary state. Any systems with dimensionless pericenters $\beta \equiv a(1-e)/R \leq  4$ should be effectively circular. In the meantime, Gaia giant binaries appear circularized out to $\beta \sim 6$ in the early RGB, and down to $\beta \sim 3$ in later stages. 
Moreover, like main-sequence binaries, Gaia giant binaries also exhibit `cool islands', over-densities of circular orbits that extend well beyond the circularization limit. These islands reach $a/R \sim 10-15$. For the early RGB, this is about twice as far as predicted by  our tidal calculations. Both of these features suggest that, while f-mode tides can adequately describe the observational data during the late RGB, there needs to be an additional circularization process during the early RGB. 

The failure of theory during the early RGB may be connected to the same failure during the main-sequence. A few candidates that could account for the excess damping have been proposed in the literature. In the future, we hope to explore these options. In addition,  we will use  Gaia binaries to study tidal processes after the RGB phase.

Lastly, we note the significant tidal spin-up experienced by many giant binaries. This could affect the giant mass-loss and affect binary evolution. Some of these binaries may also enter Roche-lobe overflow while still eccentric.

\ack{
We thank the anonymous reviewer for comments that improved the quality of the paper. We acknowledge the support of the Natural Sciences and Engineering Research Council of Canada (NSERC), [funding reference number 513671]. This work has also made use of data from the European Space Agency (ESA) mission Gaia  (https://www.cosmos.esa.int/gaia), processed by the Gaia Data Processing and Analysis Consortium (DPAC). 
}

%



\software{
numpy \citep{numpy}, 
matplotlib \citep{matplotlib}, 
scipy \citep{2020SciPy-NMeth}, 
MESA \citep{Paxton2011, Paxton2013, Paxton2015, Paxton2018, Paxton2019, Jermyn2023},
GYRE \citep{Townsend2013}
}



\appendix

\section{Equilibrium tides}\label{app:eqm}
In the limit of vanishing tidal frequency, Equations \eqref{eq:eom1}-\eqref{eq:poi1} can be combined to produce \citep{Ogilvie2014}
\begin{equation}\label{eq:eqmPhi}
    \nabla^2\delta\Phi +\frac{4\pi G\rho_0}{c_s^2}(\delta\Phi + U)=0.
\end{equation}
Together with regularity boundary conditions, and a given profile for $\rho_0$ and $c_s$, this equation can be solved numerically for the gravitational perturbation $\delta\Phi.$ In stably stratified regions, the Lagrangian displacement $\boldsymbol{\xi}$ associated with the so-called `equilibrium' tide can then be determined from
\begin{equation}\label{eq:eqmxi}
    \boldsymbol{\xi}\cdot\boldsymbol{\mathcal{G}}
    =\delta\Phi + U,
\end{equation}
together with $\nabla\cdot\boldsymbol{\xi}=0.$ This divergence-free constraint does not technically hold in regions that are isentropic \citep{Terquem1998}, and so Equation \ref{eq:eqmxi} may not adequately determine $\boldsymbol{\xi}$ in convective regions. As a work-around, \citet{Ogilvie2013} suggested that the displacement of a `non-wavelike' tide can be determined instead from the ansatz $\boldsymbol{\xi}=\nabla X$ for some scalar field $X.$ In such a case, the continuity equation 
implies that $X$ can be determined from 
\begin{equation}\label{eq:Xeqn}
    \nabla\cdot(\rho_0\nabla X)=\frac{\rho_0}{c_s^2}(\delta\Phi + U),
\end{equation}
with the boundary condition that the normal component of $\boldsymbol{\xi}$ vanish at the origin (or an interior solid boundary), and that Equation \eqref{eq:eqmxi} be satisfied at a stably stratified/isentropic interface or at the tidally perturbed star's surface. 

However, the assumption that the displacement can be determined from the gradient of a potential ignores any vortical contribution to the non-wavelike tide, and we note that the convective regions of stars are only approximately isentropic. 
Stellar models incorporating convective energy transport in fact involve a non-zero (negative) buoyancy frequency in convective regions, which in turn allows the assumption of an incompressible equilibrium tide satisfying Equation \eqref{eq:eqmxi}. 

This ambiguity motivated us to consider both equilibrium and non-wavelike calculations of $\boldsymbol{\xi}$. We find that the non-wavelike calculation generically predicts marginally weaker tidal dissipation than equilibrium tides at all times and frequencies. Since this marginal difference only leads to small differences in populations like those shown in Fig. \ref{fig:evpsnap}, we focus on equilibrium tides in this paper. 

\section{f-mode tidal calculations}\label{app:dyn}
Ignoring the Coriolis force, the linearized Equations \eqref{eq:eom1}-\eqref{eq:poi1} can be combined to form
\begin{align}\label{eq:eom2}
    \frac{\partial^2\boldsymbol{\xi}}{\partial t^2}
    -\nabla(\boldsymbol{\mathcal{G}}\cdot\boldsymbol{\xi})
    -\nabla(
        c_A^2\nabla\cdot\boldsymbol{\xi}
    )
    +\boldsymbol{\mathcal{A}}\nabla\cdot\boldsymbol{\xi}
    +\nabla\delta\Phi
    -\frac{1}{\rho_0}\nabla\cdot(2\rho_0\nu_c{\bf S})
    &=-\nabla U,
\\\label{eq:poi2}
    4\pi G \nabla\cdot(\rho_0\boldsymbol{\xi})
    +\nabla^2\delta\Phi 
    &=0,
\end{align}
where $\boldsymbol{\mathcal{A}}=\boldsymbol{\mathcal{G}}-c_A^2\nabla\ln\rho_0.$ This system is supplemented by the boundary conditions that $\boldsymbol{\xi}$ and $\delta\Phi$ remain regular at the center of the star, that the gravitational potential vanish at infinite distance, and that the normal and that the normal and tangential stresses vanish at the stellar surface. 

Inserting potentials $U=\Re\{A(r/R)^2Y_{2m}\exp[-{\rm i}\omega_{mk} t]\}$ of nominal amplitude $A$, we  compute dynamical ($\omega=\omega_{mk}\not=0$) solutions $\boldsymbol{\xi}_{mk}$ with the time dependence $\exp[-\text{i}\omega_{mk}t]$ to Equations \ref{eq:eom2}-\ref{eq:poi2} through an expansion in eigenmodes of the homogeneous problem (with $U=0$): $\boldsymbol{\xi}_{mk}=\sum_\alpha \tilde{a}_\alpha\boldsymbol{\xi}_\alpha,$ where $\alpha$ labels oscillation modes with displacement eigenfunctions $\boldsymbol{\xi}_\alpha$ and natural frequencies $\omega_\alpha$. The amplitudes $\tilde{a}_\alpha(t)=a_\alpha\exp[-{\rm i}\omega_{mk}t]$ can then be determined approximately from \citep[e.g.,][]{Schenk2001}
\begin{equation}\label{eq:amp}
    a_\alpha\approx-(\omega_\alpha^2 - \omega_{mk}^2 - 2\text{i}\gamma_\alpha\omega_{mk})^{-1}
    AQ_\alpha,
\end{equation}
where $Q_\alpha=\int_{V}\rho_0\boldsymbol{\xi}_\alpha^*\cdot\nabla[(r/R)^2Y_{2m}]\text{d}V$, and $\gamma_\alpha$ gives the damping rate a given mode $\alpha$ would have if it were freely oscillating in isolation. Since many oscillations are driven simultaneously rather than in isolation,  these `linear' damping rates do not adequately describe viscous dissipation \citep{Braviner2015}. However, their inclusion in Equation \ref{eq:amp} regularizes resonances (where $\omega_\alpha=\omega_{mk}$). We then compute the amplitude-normalized dissipation associated with tidal driving at the frequency $\omega_{mk}=k\Omega_o-m\Omega$ as
\begin{equation}\label{eq:Dmode}
    \mathcal{D}_{m k}
    =\frac{\omega_{mk}^2}{|A|^2}\sum_{\alpha,\beta}
    \left(\frac{a_\alpha^*a_\beta}{\omega_\alpha\omega_\beta}\right)
    \int_{V}\rho_0\nu_c{\bf S}_\alpha^*:{\bf S}_\beta\text{d}V,
\end{equation}
where $\beta$ is another mode label, and ${\bf S}_\alpha$ is the shear tensor (see Equation \ref{eq:SS}) involving the velocity eigenfunction ${\bf v}_\alpha=-{\rm i}\omega_\alpha\boldsymbol{\xi}_\alpha$ of the mode $\alpha$. Viscous coupling between modes with $\alpha\not=\beta$ affects the dissipation rate only marginally in this case. However, the interrelated frequency and spatial dependence of the effective viscosity (see Equation \ref{eq:numod}) is \emph{essential}. This frequency dependence cannot be captured by the isolated damping rates $\gamma_\alpha,$ instead necessitating on-the-fly calculation of the integrals appearing in Equation \eqref{eq:Dmode}.

We use a mode expansion that includes the degree $2$ f-mode and the ten lowest order p-modes. We pre-compute these oscillations for each MESA timestep with GYRE \citep{Townsend2013}. Because we focus on the later stages of the RGB, we place the inner boundary at the outer edge of the radiative core, and adopt \texttt{GYRE}'s \texttt{ZERO\_H} inner boundary condition. We linearly interpolate the stellar profiles  
($\rho_0,$ 
$\ell_c$, 
$v_c$) 
and oscillation mode properties 
($\omega_\alpha$, 
$\gamma_\alpha,$ 
$Q_\alpha$, 
${\bf S}_\alpha^*:{\bf S}_\beta$) 
required to compute tidal dissipation rates at each step of our integration of the secular tidal equations \eqref{eq:adot}-\eqref{eq:Jdot}. 

\section{Tidal energy and angular momentum transfer}\label{app:EAM}
In the co-rotating, co-moving frame of a tidally perturbed star, the total tidal potential imposed by a point mass, spin-aligned perturber can be written as
\begin{equation}\label{eq:Uexp}
    U=\sum_{n=2}^\infty
    \sum_{m=-n}^n
    \sum_{k=-\infty}^\infty
    U_{nmk}\left(\frac{r}{R}\right)^n
    Y_{nm}
    e^{-\text{i}\omega_{mk}t},
\end{equation}
where $Y_{nm}(\theta,\phi)$ are ortho-normalized spherical harmonics, $\omega_{mk}=k\Omega_o-m\Omega,$ and 
\begin{equation}
    U_{nmk}
    =-\dfrac{GM'}{a}
    \left(\frac{R}{a}\right)^n
    \frac{4\pi }{(2n+1)}Y_{nm}^{*}(\pi/2,0)X_{nmk},
\end{equation}
where $X_{nmk}$ are Hansen coefficients \citep[see, e.g., eq. 15 in][]{Zanazzi2021}. The degree $n=0$ part of a point-mass potential has been omitted from Equation \eqref{eq:Uexp} because it does not produce a force, while the $n=1$ part cancels in the frame co-moving with the primary.

Fluid motions induced by $U$ give rise to a potential perturbation $\delta \Phi$ that in the external vacuum can be written in terms of the expansion
\begin{equation}
    \delta\Phi 
    =\sum_{\ell=0}^\infty
    \sum_{m=-\ell}^\ell
    \sum_{k=-\infty}^\infty
    \delta\Phi_{\ell m}^k\left(\frac{R}{r}\right)^{\ell+1}
    Y_{\ell m}
    e^{-\text{i}\omega_{mk}t}.
\end{equation}
In an axisymmetric star, tidal Love numbers can be defined to satisfy the linear relation $\delta\Phi_{\ell m}^k=\sum_n k_{\ell m}^nU_{nmk},$ where $k_{\ell m}^n$ are functions of tidal frequency \citep{Ogilvie2013}. For a spherically symmetric (i.e., non-rotating) star, $k_{\ell m}^n=0$ unless $n=\ell$.

Truncating at $n=2$, we suppress subscripts $n$. In a non-rotating frame co-moving with the primary star, time-steady rates of energy and angular momentum transfer can be found as 
\begin{align}
    \mathcal{P}
    &=\frac{5R}{4\pi G}
    \sum_{m=-2}^2
    \sum_{k=-\infty}^\infty
    k\Omega_o\kappa_{mk}|U_{mk}|^2,
\\
    \mathcal{T}
    &=\frac{5R}{4\pi G}
    \sum_{m=-2}^2
    \sum_{k=-\infty}^\infty
    m\kappa_{mk}|U_{mk}|^2.
\end{align}
Here $\kappa_{mk}=\Im[k_{2m}^2]$ are the imaginary parts of $\ell=n=2$ tidal Love numbers. These can can be determined from Equations \eqref{eq:Dmode} and \eqref{eq:Dtokap} for a given $m$ and $k.$ Then, equating time derivatives of the orbital energy $E_o=-GMM'/(2a)$, orbital angular momentum $L_o=-2(1-e^2)^{1/2}E_o/\Omega_o$, and stellar angular momentum $J$ with the rates of energy and angular momentum transfer ($\mathcal{P}$, $-\mathcal{T}$, and $\mathcal{T},$ respectively) leads to the evolutionary Equations \eqref{eq:adot}-\eqref{eq:Jdot}.

\section{Orbital integration sample}\label{app:samp}
\begin{figure*}
    \centering
    \includegraphics[width=0.9\textwidth]{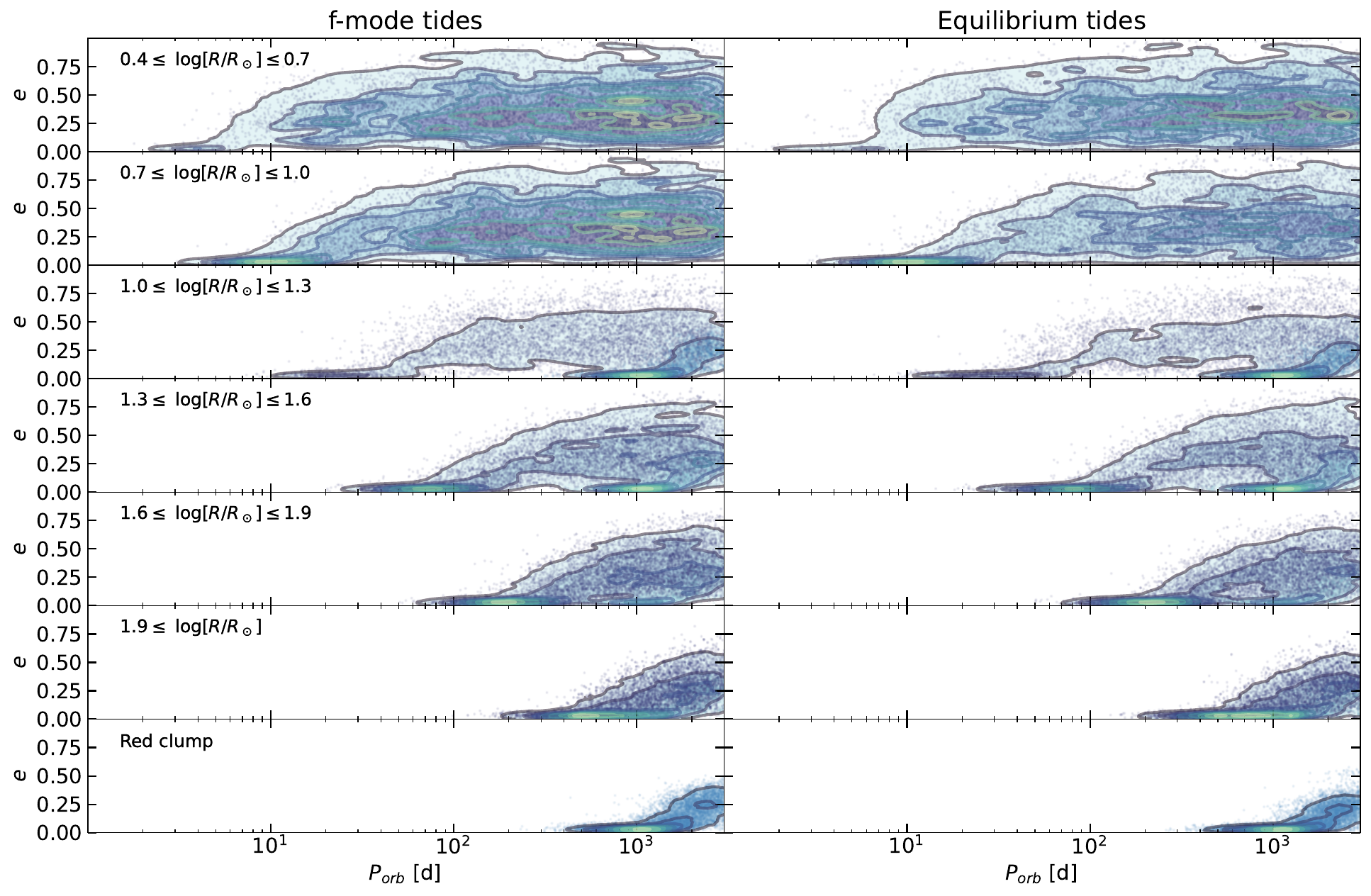}
    \caption{Scatterplots and kernel density estimates showing eccentricity and orbital period values, sampled from orbital evolution calculations sliced evenly in time (point colors again indicate evolutionary stage), and perturbed as described in Appendix \ref{app:samp}. The contours indicate normalized KDE levels $\sim[0.05,0.18,...,1]$. With increasing age and radius for the primary star, tidal circularization extends to wider and wider orbital periods. }\label{fig:evpsnap}
\end{figure*}

The scatterplots and kernel density estimate contours in Fig. \ref{fig:evpsnap} show data obtained by integrating the tidal equations from the initial conditions described in Section \ref{sec:ics}. After translating $P_{\rm orb}$ to $a/R$, we use these data to generate the kernel density estimates shown in Fig. \ref{fig:fmode_obs_cpr} (right). 

We integrate each initial condition until the onset of common envelope evolution (i.e., until $(1-e)a\leq R$), assuming that RLOF does not significantly affect tidal evolution (simply halting evolution at RLOF produces similar results). To sample our evolutionary tracks, we first average computed values of $e$ and $P_\text{orb}$ within evenly spaced time increments of $10^5$ years, sorting by the same radial bins shown in Fig. \ref{fig:hrd_gaia} and Fig. \ref{fig:RM} (middle). To each averaged period $P_\text{orb}$, we then add a perturbation drawn from a normal distribution with a standard deviation equal to $0.2P_\text{orb}.$ Assuming a random phase, we separate each averaged eccentricity amplitude into $x$ and $y$ components of an eccentricity vector, and then perturb each vector component with errors drawn from a normal distribution with a standard deviation of $0.025$. These perturbations are only added to scatter a number of points that would otherwise lie on top of each other, and do not significantly change our results.

\bibliography{RGBtides}{}

\begin{thebibliography}{}
\expandafter\ifx\csname natexlab\endcsname\relax\def\natexlab#1{#1}\fi
\providecommand{\url}[1]{\href{#1}{#1}}
\providecommand{\dodoi}[1]{doi:~\href{http://doi.org/#1}{\nolinkurl{#1}}}
\providecommand{\doeprint}[1]{\href{http://ascl.net/#1}{\nolinkurl{http://ascl.net/#1}}}
\providecommand{\doarXiv}[1]{\href{https://arxiv.org/abs/#1}{\nolinkurl{https://arxiv.org/abs/#1}}}

\bibitem[{{Barker}(2022)}]{Barker2022}
{Barker}, A.~J. 2022, \apjl, 927, L36, \dodoi{10.3847/2041-8213/ac5b63}

\bibitem[{{Barker} \& {Astoul}(2021)}]{Barker2021}
{Barker}, A.~J., \& {Astoul}, A. A.~V. 2021, \mnras, 506, L69, \dodoi{10.1093/mnrasl/slab077}

\bibitem[{{Bashi} {et~al.}(2023){Bashi}, {Mazeh}, \& {Faigler}}]{Bashi2023}
{Bashi}, D., {Mazeh}, T., \& {Faigler}, S. 2023, \mnras, 522, 1184, \dodoi{10.1093/mnras/stad999}

\bibitem[{{Beck} {et~al.}(2018){Beck}, {Mathis}, {Gallet}, {Charbonnel}, {Benbakoura}, {Garc{\'\i}a}, \& {do Nascimento}}]{Beck2018}
{Beck}, P.~G., {Mathis}, S., {Gallet}, F., {et~al.} 2018, \mnras, 479, L123, \dodoi{10.1093/mnrasl/sly114}

\bibitem[{{Beck} {et~al.}(2024){Beck}, {Grossmann}, {Steinwender}, {Schimak}, {Muntean}, {Vrard}, {Patton}, {Merc}, {Mathur}, {Garcia}, {Pinsonneault}, {Rowan}, {Gaulme}, {Allende Prieto}, {Arellano-C{\'o}rdova}, {Cao}, {Corsaro}, {Creevey}, {Hambleton}, {Hanslmeier}, {Holl}, {Johnson}, {Mathis}, {Godoy-Rivera}, {S{\'\i}mon-D{\'\i}az}, \& {Zinn}}]{Beck2024}
{Beck}, P.~G., {Grossmann}, D.~H., {Steinwender}, L., {et~al.} 2024, \aap, 682, A7, \dodoi{10.1051/0004-6361/202346810}

\bibitem[{{Bl\"ocker}(1995)}]{Bloecker1995}
{Bl\"ocker}, T. 1995, \aap, 297, 727

\bibitem[{{Braviner} \& {Ogilvie}(2015)}]{Braviner2015}
{Braviner}, H.~J., \& {Ogilvie}, G.~I. 2015, \mnras, 447, 1141, \dodoi{10.1093/mnras/stu2521}

\bibitem[{{Bressan} {et~al.}(2012){Bressan}, {Marigo}, {Girardi}, {Salasnich}, {Dal Cero}, {Rubele}, \& {Nanni}}]{parsec}
{Bressan}, A., {Marigo}, P., {Girardi}, L., {et~al.} 2012, \mnras, 427, 127, \dodoi{10.1111/j.1365-2966.2012.21948.x}

\bibitem[{{Bryan} {et~al.}(2024){Bryan}, {de Wit}, {Sun}, {de Beurs}, \& {Townsend}}]{Bryan2024}
{Bryan}, J., {de Wit}, J., {Sun}, M., {de Beurs}, Z.~L., \& {Townsend}, R. H.~D. 2024, Nature Astronomy, 8, 1387, \dodoi{10.1038/s41550-024-02351-3}

\bibitem[{{Duguid} {et~al.}(2020){Duguid}, {Barker}, \& {Jones}}]{Duguid2020}
{Duguid}, C.~D., {Barker}, A.~J., \& {Jones}, C.~A. 2020, \mnras, 497, 3400, \dodoi{10.1093/mnras/staa2216}

\bibitem[{{Duguid} {et~al.}(2024){Duguid}, {de Vries}, {Lecoanet}, \& {Barker}}]{Duguid2024}
{Duguid}, C.~D., {de Vries}, N.~B., {Lecoanet}, D., \& {Barker}, A.~J. 2024, \apjl, 966, L14, \dodoi{10.3847/2041-8213/ad3c40}

\bibitem[{{Eggleton}(1983)}]{Eggleton1983}
{Eggleton}, P.~P. 1983, \apj, 268, 368, \dodoi{10.1086/160960}

\bibitem[{{Esseldeurs} {et~al.}(2024){Esseldeurs}, {Mathis}, \& {Decin}}]{Esseldeurs2024}
{Esseldeurs}, M., {Mathis}, S., \& {Decin}, L. 2024, arXiv e-prints, arXiv:2407.10573, \dodoi{10.48550/arXiv.2407.10573}

\bibitem[{{Fuller}(2017)}]{Fuller2017}
{Fuller}, J. 2017, \mnras, 472, 1538, \dodoi{10.1093/mnras/stx2135}

\bibitem[{{Gaia Collaboration} {et~al.}(2023){Gaia Collaboration}, {Arenou}, {Babusiaux}, {Barstow}, {Faigler}, {Jorissen}, {Kervella}, {Mazeh}, {Mowlavi}, {Panuzzo}, {Sahlmann}, {Shahaf}, {Sozzetti}, {Bauchet}, {Damerdji}, {Gavras}, {Giacobbe}, {Gosset}, {Halbwachs}, {Holl}, {Lattanzi}, {Leclerc}, {Morel}, {Pourbaix}, {Re Fiorentin}, {Sadowski}, {S{\'e}gransan}, {Siopis}, {Teyssier}, {Zwitter}, {Planquart}, {Brown}, {Vallenari}, {Prusti}, {de Bruijne}, {Biermann}, {Creevey}, {Ducourant}, {Evans}, {Eyer}, {Guerra}, {Hutton}, {Jordi}, {Klioner}, {Lammers}, {Lindegren}, {Luri}, {Mignard}, {Panem}, {Randich}, {Sartoretti}, {Soubiran}, {Tanga}, {Walton}, {Bailer-Jones}, {Bastian}, {Drimmel}, {Jansen}, {Katz}, {van Leeuwen}, {Bakker}, {Cacciari}, {Casta{\~n}eda}, {De Angeli}, {Fabricius}, {Fouesneau}, {Fr{\'e}mat}, {Galluccio}, {Guerrier}, {Heiter}, {Masana}, {Messineo}, {Nicolas}, {Nienartowicz}, {Pailler}, {Riclet}, {Roux}, {Seabroke}, {Sordo}, {Th{\'e}venin}, {Gracia-Abril}, {Portell}, {Altmann}, {Andrae},
  {Audard}, {Bellas-Velidis}, {Benson}, {Berthier}, {Blomme}, {Burgess}, {Busonero}, {Busso}, {C{\'a}novas}, {Carry}, {Cellino}, {Cheek}, {Clementini}, {Davidson}, {de Teodoro}, {Nu{\~n}ez Campos}, {Delchambre}, {Dell'Oro}, {Esquej}, {Fern{\'a}ndez-Hern{\'a}ndez}, {Fraile}, {Garabato}, {Garc{\'\i}a-Lario}, {Haigron}, {Hambly}, {Harrison}, {Hern{\'a}ndez}, {Hestroffer}, {Hodgkin}, {Jan{\ss}en}, {Jevardat de Fombelle}, {Jordan}, {Krone-Martins}, {Lanzafame}, {L{\"o}ffler}, {Marchal}, {Marrese}, {Moitinho}, {Muinonen}, {Osborne}, {Pancino}, {Pauwels}, {Recio-Blanco}, {Reyl{\'e}}, {Riello}, {Rimoldini}, {Roegiers}, {Rybizki}, {Sarro}, {Smith}, {Utrilla}, {van Leeuwen}, {Abbas}, {{\'A}brah{\'a}m}, {Abreu Aramburu}, {Aerts}, {Aguado}, {Ajaj}, {Aldea-Montero}, {Altavilla}, {{\'A}lvarez}, {Alves}, {Anders}, {Anderson}, {Anglada Varela}, {Antoja}, {Baines}, {Baker}, {Balaguer-N{\'u}{\~n}ez}, {Balbinot}, {Balog}, {Barache}, {Barbato}, {Barros}, {Bartolom{\'e}}, {Bassilana}, {Becciani}, {Bellazzini}, {Berihuete},
  {Bernet}, {Bertone}, {Bianchi}, {Binnenfeld}, {Blanco-Cuaresma}, {Blazere}, {Boch}, {Bombrun}, {Bossini}, {Bouquillon}, {Bragaglia}, {Bramante}, {Breedt}, {Bressan}, {Brouillet}, {Brugaletta}, {Bucciarelli}, {Burlacu}, {Butkevich}, {Buzzi}, {Caffau}, {Cancelliere}, {Cantat-Gaudin}, {Carballo}, {Carlucci}, {Carnerero}, {Carrasco}, {Casamiquela}, {Castellani}, {Castro-Ginard}, {Chaoul}, {Charlot}, {Chemin}, {Chiaramida}, {Chiavassa}, {Chornay}, {Comoretto}, {Contursi}, {Cooper}, {Cornez}, {Cowell}, {Crifo}, {Cropper}, {Crosta}, {Crowley}, {Dafonte}, {Dapergolas}, {David}, {de Laverny}, {De Luise}, {De March}, {De Ridder}, {de Souza}, {de Torres}, {del Peloso}, {del Pozo}, {Delbo}, {Delgado}, {Delisle}, {Demouchy}, {Dharmawardena}, {Diakite}, {Diener}, {Distefano}, {Dolding}, {Enke}, {Fabre}, {Fabrizio}, {Fedorets}, {Fernique}, {Figueras}, {Fournier}, {Fouron}, {Fragkoudi}, {Gai}, {Garcia-Gutierrez}, {Garcia-Reinaldos}, {Garc{\'\i}a-Torres}, {Garofalo}, {Gavel}, {Gerlach}, {Geyer}, {Gilmore}, {Girona},
  {Giuffrida}, {Gomel}, {Gomez}, {Gonz{\'a}lez-N{\'u}{\~n}ez}, {Gonz{\'a}lez-Santamar{\'\i}a}, {Gonz{\'a}lez-Vidal}, {Granvik}, {Guillout}, {Guiraud}, {Guti{\'e}rrez-S{\'a}nchez}, {Guy}, {Hatzidimitriou}, {Hauser}, {Haywood}, {Helmer}, {Helmi}, {Sarmiento}, {Hidalgo}, {Hilger}, {H{\l}adczuk}, {Hobbs}, {Holland}, {Huckle}, {Jardine}, {Jasniewicz}, {Jean-Antoine Piccolo}, {Jim{\'e}nez-Arranz}, {Juaristi Campillo}, {Julbe}, {Karbevska}, {Khanna}, {Kordopatis}, {Korn}, {K{\'o}sp{\'a}l}, {Kostrzewa-Rutkowska}, {Kruszy{\'n}ska}, {Kun}, {Laizeau}, {Lambert}, {Lanza}, {Lasne}, {Le Campion}, {Lebreton}, {Lebzelter}, {Leccia}, {Lecoeur-Taibi}, {Liao}, {Licata}, {Lindstr{\o}m}, {Lister}, {Livanou}, {Lobel}, {Lorca}, {Loup}, {Madrero Pardo}, {Magdaleno Romeo}, {Managau}, {Mann}, {Manteiga}, {Marchant}, {Marconi}, {Marcos}, {Marcos Santos}, {Mar{\'\i}n Pina}, {Marinoni}, {Marocco}, {Marshall}, {Martin Polo}, {Mart{\'\i}n-Fleitas}, {Marton}, {Mary}, {Masip}, {Massari}, {Mastrobuono-Battisti}, {McMillan}, {Messina},
  {Michalik}, {Millar}, {Mints}, {Molina}, {Molinaro}, {Moln{\'a}r}, {Monari}, {Mongui{\'o}}, {Montegriffo}, {Montero}, {Mor}, {Mora}, {Morbidelli}, {Morris}, {Muraveva}, {Murphy}, {Musella}, {Nagy}, {Noval}, {Oca{\~n}a}, {Ogden}, {Ordenovic}, {Osinde}, {Pagani}, {Pagano}, {Palaversa}, {Palicio}, {Pallas-Quintela}, {Panahi}, {Payne-Wardenaar}, {Pe{\~n}alosa Esteller}, {Penttil{\"a}}, {Pichon}, {Piersimoni}, {Pineau}, {Plachy}, {Plum}, {Poggio}, {Pr{\v{s}}a}, {Pulone}, {Racero}, {Ragaini}, {Rainer}, {Raiteri}, {Ramos}, {Ramos-Lerate}, {Regibo}, {Richards}, {Rios Diaz}, {Ripepi}, {Riva}, {Rix}, {Rixon}, {Robichon}, {Robin}, {Robin}, {Roelens}, {Rogues}, {Rohrbasser}, {Romero-G{\'o}mez}, {Rowell}, {Royer}, {Ruz Mieres}, {Rybicki}, {S{\'a}ez N{\'u}{\~n}ez}, {Sagrist{\`a} Sell{\'e}s}, {Salguero}, {Samaras}, {Sanchez Gimenez}, {Sanna}, {Santove{\~n}a}, {Sarasso}, {Schultheis}, {Sciacca}, {Segol}, {Segovia}, {Semeux}, {Siddiqui}, {Siebert}, {Siltala}, {Silvelo}, {Slezak}, {Slezak}, {Smart}, {Snaith}, {Solano},
  {Solitro}, {Souami}, {Souchay}, {Spagna}, {Spina}, {Spoto}, {Steele}, {Steidelm{\"u}ller}, {Stephenson}, {S{\"u}veges}, {Surdej}, {Szabados}, {Szegedi-Elek}, {Taris}, {Taylor}, {Teixeira}, {Tolomei}, {Tonello}, {Torra}, {Torra}, {Torralba Elipe}, {Trabucchi}, {Tsounis}, {Turon}, {Ulla}, {Unger}, {Vaillant}, {van Dillen}, {van Reeven}, {Vanel}, {Vecchiato}, {Viala}, {Vicente}, {Voutsinas}, {Weiler}, {Wevers}, {Wyrzykowski}, {Yoldas}, {Yvard}, {Zhao}, {Zorec}, \& {Zucker}}]{Arenou2023}
{Gaia Collaboration}, {Arenou}, F., {Babusiaux}, C., {et~al.} 2023, \aap, 674, A34, \dodoi{10.1051/0004-6361/202243782}

\bibitem[{{Goldreich} \& {Nicholson}(1977)}]{Goldreich1977}
{Goldreich}, P., \& {Nicholson}, P.~D. 1977, \icarus, 30, 301, \dodoi{10.1016/0019-1035(77)90163-4}

\bibitem[{{Goodman} \& {Dickson}(1998)}]{Goodman1998}
{Goodman}, J., \& {Dickson}, E.~S. 1998, \apj, 507, 938, \dodoi{10.1086/306348}

\bibitem[{{Goodman} \& {Oh}(1997)}]{Goodman1997}
{Goodman}, J., \& {Oh}, S.~P. 1997, \apj, 486, 403, \dodoi{10.1086/304505}

\bibitem[{{Guo} {et~al.}(2023){Guo}, {Ogilvie}, \& {Barker}}]{Guo2023}
{Guo}, Z., {Ogilvie}, G.~I., \& {Barker}, A.~J. 2023, \mnras, 521, 1353, \dodoi{10.1093/mnras/stad569}

\bibitem[{{Han} {et~al.}(2020){Han}, {Ge}, {Chen}, \& {Chen}}]{Han2020}
{Han}, Z.-W., {Ge}, H.-W., {Chen}, X.-F., \& {Chen}, H.-L. 2020, Research in Astronomy and Astrophysics, 20, 161, \dodoi{10.1088/1674-4527/20/10/161}

\bibitem[{Harris {et~al.}(2020)Harris, Millman, van~der Walt, Gommers, Virtanen, Cournapeau, Wieser, Taylor, Berg, Smith, Kern, Picus, Hoyer, van Kerkwijk, Brett, Haldane, del R{\'{i}}o, Wiebe, Peterson, G{\'{e}}rard-Marchant, Sheppard, Reddy, Weckesser, Abbasi, Gohlke, \& Oliphant}]{numpy}
Harris, C.~R., Millman, K.~J., van~der Walt, S.~J., {et~al.} 2020, Nature, 585, 357, \dodoi{10.1038/s41586-020-2649-2}

\bibitem[{Hunter(2007)}]{matplotlib}
Hunter, J.~D. 2007, Computing in Science \& Engineering, 9, 90, \dodoi{10.1109/MCSE.2007.55}

\bibitem[{{Hurley} {et~al.}(2002){Hurley}, {Tout}, \& {Pols}}]{Hurley2002}
{Hurley}, J.~R., {Tout}, C.~A., \& {Pols}, O.~R. 2002, \mnras, 329, 897, \dodoi{10.1046/j.1365-8711.2002.05038.x}

\bibitem[{{Jermyn} {et~al.}(2023){Jermyn}, {Bauer}, {Schwab}, {Farmer}, {Ball}, {Bellinger}, {Dotter}, {Joyce}, {Marchant}, {Mombarg}, {Wolf}, {Sunny Wong}, {Cinquegrana}, {Farrell}, {Smolec}, {Thoul}, {Cantiello}, {Herwig}, {Toloza}, {Bildsten}, {Townsend}, \& {Timmes}}]{Jermyn2023}
{Jermyn}, A.~S., {Bauer}, E.~B., {Schwab}, J., {et~al.} 2023, \apjs, 265, 15, \dodoi{10.3847/1538-4365/acae8d}

\bibitem[{{Leconte} {et~al.}(2010){Leconte}, {Chabrier}, {Baraffe}, \& {Levrard}}]{Leconte2010}
{Leconte}, J., {Chabrier}, G., {Baraffe}, I., \& {Levrard}, B. 2010, \aap, 516, A64, \dodoi{10.1051/0004-6361/201014337}

\bibitem[{{Lin} \& {Ogilvie}(2018)}]{Lin2018}
{Lin}, Y., \& {Ogilvie}, G.~I. 2018, \mnras, 474, 1644, \dodoi{10.1093/mnras/stx2764}

\bibitem[{{Mazeh}(2008)}]{Mazeh2008}
{Mazeh}, T. 2008, in EAS Publications Series, Vol.~29, Tidal Effects in Stars, Planets and Disks, ed. M.~J. {Goupil} \& J.~P. {Zahn}, 1--65, \dodoi{10.1051/eas:0829001}

\bibitem[{{Meibom} \& {Mathieu}(2005)}]{Meibom2005}
{Meibom}, S., \& {Mathieu}, R.~D. 2005, \apj, 620, 970, \dodoi{10.1086/427082}

\bibitem[{{Meibom} {et~al.}(2006){Meibom}, {Mathieu}, \& {Stassun}}]{Meibom2006}
{Meibom}, S., {Mathieu}, R.~D., \& {Stassun}, K.~G. 2006, \apj, 653, 621, \dodoi{10.1086/508252}

\bibitem[{{Moe} \& {Di Stefano}(2017)}]{Moe2017}
{Moe}, M., \& {Di Stefano}, R. 2017, \apjs, 230, 15, \dodoi{10.3847/1538-4365/aa6fb6}

\bibitem[{{Mowlavi} {et~al.}(2023){Mowlavi}, {Holl}, {Lecoeur-Ta{\"\i}bi}, {Barblan}, {Kochoska}, {Pr{\v{s}}a}, {Mazeh}, {Rimoldini}, {Gavras}, {Audard}, {Jevardat de Fombelle}, {Nienartowicz}, {Garc{\'\i}a-Lario}, \& {Eyer}}]{Mowlavi2023}
{Mowlavi}, N., {Holl}, B., {Lecoeur-Ta{\"\i}bi}, I., {et~al.} 2023, \aap, 674, A16, \dodoi{10.1051/0004-6361/202245330}

\bibitem[{{Ogilvie}(2013)}]{Ogilvie2013}
{Ogilvie}, G.~I. 2013, \mnras, 429, 613, \dodoi{10.1093/mnras/sts362}

\bibitem[{{Ogilvie}(2014)}]{Ogilvie2014}
---. 2014, \araa, 52, 171, \dodoi{10.1146/annurev-astro-081913-035941}

\bibitem[{{Ogilvie} \& {Lin}(2007)}]{Ogilvie2007}
{Ogilvie}, G.~I., \& {Lin}, D.~N.~C. 2007, \apj, 661, 1180, \dodoi{10.1086/515435}

\bibitem[{{Paxton} {et~al.}(2011){Paxton}, {Bildsten}, {Dotter}, {Herwig}, {Lesaffre}, \& {Timmes}}]{Paxton2011}
{Paxton}, B., {Bildsten}, L., {Dotter}, A., {et~al.} 2011, \apjs, 192, 3, \dodoi{10.1088/0067-0049/192/1/3}

\bibitem[{{Paxton} {et~al.}(2013){Paxton}, {Cantiello}, {Arras}, {Bildsten}, {Brown}, {Dotter}, {Mankovich}, {Montgomery}, {Stello}, {Timmes}, \& {Townsend}}]{Paxton2013}
{Paxton}, B., {Cantiello}, M., {Arras}, P., {et~al.} 2013, \apjs, 208, 4, \dodoi{10.1088/0067-0049/208/1/4}

\bibitem[{{Paxton} {et~al.}(2015){Paxton}, {Marchant}, {Schwab}, {Bauer}, {Bildsten}, {Cantiello}, {Dessart}, {Farmer}, {Hu}, {Langer}, {Townsend}, {Townsley}, \& {Timmes}}]{Paxton2015}
{Paxton}, B., {Marchant}, P., {Schwab}, J., {et~al.} 2015, \apjs, 220, 15, \dodoi{10.1088/0067-0049/220/1/15}

\bibitem[{{Paxton} {et~al.}(2018){Paxton}, {Schwab}, {Bauer}, {Bildsten}, {Blinnikov}, {Duffell}, {Farmer}, {Goldberg}, {Marchant}, {Sorokina}, {Thoul}, {Townsend}, \& {Timmes}}]{Paxton2018}
{Paxton}, B., {Schwab}, J., {Bauer}, E.~B., {et~al.} 2018, \apjs, 234, 34, \dodoi{10.3847/1538-4365/aaa5a8}

\bibitem[{{Paxton} {et~al.}(2019){Paxton}, {Smolec}, {Schwab}, {Gautschy}, {Bildsten}, {Cantiello}, {Dotter}, {Farmer}, {Goldberg}, {Jermyn}, {Kanbur}, {Marchant}, {Thoul}, {Townsend}, {Wolf}, {Zhang}, \& {Timmes}}]{Paxton2019}
{Paxton}, B., {Smolec}, R., {Schwab}, J., {et~al.} 2019, \apjs, 243, 10, \dodoi{10.3847/1538-4365/ab2241}

\bibitem[{{Pichon}(2007)}]{flame2007}
{Pichon}, B. 2007, in SF2A-2007: Proceedings of the Annual meeting of the French Society of Astronomy and Astrophysics, ed. J.~{Bouvier}, A.~{Chalabaev}, \& C.~{Charbonnel}, 549

\bibitem[{{Price-Whelan} \& {Goodman}(2018)}]{Price-Whelan2018}
{Price-Whelan}, A.~M., \& {Goodman}, J. 2018, \apj, 867, 5, \dodoi{10.3847/1538-4357/aae264}

\bibitem[{{Rafikov}(2016)}]{Rafikov2016}
{Rafikov}, R.~R. 2016, \apj, 830, 8, \dodoi{10.3847/0004-637X/830/1/8}

\bibitem[{Reimers(1975)}]{Reimers1975}
Reimers, D. 1975, Circumstellar Envelopes and Mass Loss of Red Giant Stars (Berlin, Heidelberg: Springer Berlin Heidelberg), 229--256, \dodoi{10.1007/978-3-642-80919-4_8}

\bibitem[{{Schenk} {et~al.}(2001){Schenk}, {Arras}, {Flanagan}, {Teukolsky}, \& {Wasserman}}]{Schenk2001}
{Schenk}, A.~K., {Arras}, P., {Flanagan}, {\'E}.~{\'E}., {Teukolsky}, S.~A., \& {Wasserman}, I. 2001, \prd, 65, 024001, \dodoi{10.1103/PhysRevD.65.024001}

\bibitem[{{Terquem}(2021)}]{Terquem2021}
{Terquem}, C. 2021, \mnras, 503, 5789, \dodoi{10.1093/mnras/stab224}

\bibitem[{{Terquem}(2023)}]{Terquem2023}
---. 2023, \mnras, 525, 508, \dodoi{10.1093/mnras/stad2163}

\bibitem[{{Terquem} {et~al.}(1998){Terquem}, {Papaloizou}, {Nelson}, \& {Lin}}]{Terquem1998}
{Terquem}, C., {Papaloizou}, J.~C.~B., {Nelson}, R.~P., \& {Lin}, D.~N.~C. 1998, \apj, 502, 788, \dodoi{10.1086/305927}

\bibitem[{{Townsend} \& {Teitler}(2013)}]{Townsend2013}
{Townsend}, R.~H.~D., \& {Teitler}, S.~A. 2013, \mnras, 435, 3406, \dodoi{10.1093/mnras/stt1533}

\bibitem[{{Verbunt} \& {Phinney}(1995)}]{Verbunt1995}
{Verbunt}, F., \& {Phinney}, E.~S. 1995, \aap, 296, 709

\bibitem[{{Vick} \& {Lai}(2020)}]{Vick2020}
{Vick}, M., \& {Lai}, D. 2020, \mnras, 496, 3767, \dodoi{10.1093/mnras/staa1784}

\bibitem[{{Vick} {et~al.}(2021){Vick}, {MacLeod}, {Lai}, \& {Loeb}}]{Vick2021}
{Vick}, M., {MacLeod}, M., {Lai}, D., \& {Loeb}, A. 2021, \mnras, 503, 5569, \dodoi{10.1093/mnras/stab850}

\bibitem[{Virtanen {et~al.}(2020)Virtanen, Gommers, Oliphant, Haberland, Reddy, Cournapeau, Burovski, Peterson, Weckesser, Bright, {van der Walt}, Brett, Wilson, Millman, Mayorov, Nelson, Jones, Kern, Larson, Carey, Polat, Feng, Moore, {VanderPlas}, Laxalde, Perktold, Cimrman, Henriksen, Quintero, Harris, Archibald, Ribeiro, Pedregosa, {van Mulbregt}, \& {SciPy 1.0 Contributors}}]{2020SciPy-NMeth}
Virtanen, P., Gommers, R., Oliphant, T.~E., {et~al.} 2020, Nature Methods, 17, 261, \dodoi{10.1038/s41592-019-0686-2}

\bibitem[{{Weinberg} {et~al.}(2017){Weinberg}, {Sun}, {Arras}, \& {Essick}}]{Weinberg2017}
{Weinberg}, N.~N., {Sun}, M., {Arras}, P., \& {Essick}, R. 2017, \apjl, 849, L11, \dodoi{10.3847/2041-8213/aa9113}

\bibitem[{{Witte} \& {Savonije}(1999)}]{Witte1999}
{Witte}, M.~G., \& {Savonije}, G.~J. 1999, \aap, 350, 129, \dodoi{10.48550/arXiv.astro-ph/9909073}

\bibitem[{{Wu} {et~al.}(2024){Wu}, {Hadden}, {Dewberry}, {El-Badry}, \& {Matzner}}]{Wu2024}
{Wu}, Y., {Hadden}, S., {Dewberry}, J., {El-Badry}, K., \& {Matzner}, C.~D. 2024, arXiv e-prints, arXiv:2411.09905, \dodoi{10.48550/arXiv.2411.09905}

\bibitem[{{Xu} \& {Lai}(2017)}]{Xu2017}
{Xu}, W., \& {Lai}, D. 2017, \prd, 96, 083005, \dodoi{10.1103/PhysRevD.96.083005}

\bibitem[{{Zahn}(1977)}]{Zahn1977}
{Zahn}, J.~P. 1977, \aap, 57, 383

\bibitem[{{Zanazzi} \& {Wu}(2021)}]{Zanazzi2021}
{Zanazzi}, J.~J., \& {Wu}, Y. 2021, \aj, 161, 263, \dodoi{10.3847/1538-3881/abf097}

\end{thebibliography}
\bibliographystyle{aasjournal}

\end{document}